# Lead-Free Europium Halide Perovskite Nanoplatelets

*Sebastian Fernández,[1,†] Divine Mbachu,[1,†] Manchen Hu,[1] Han Cui,[2,3] William Michaels,[1] Pournima Narayanan,[1,4] Tyler K. Colenbrander,[1] Qi Zhou,[1] Da Lin,[1,2] Ona Segura Lecina,[1] Guosong Hong,[2,3] Daniel N. Congreve*,[1]*

1: Department of Electrical Engineering, Stanford University, Stanford, CA, USA
2: Department of Materials Science and Engineering, Stanford University, Stanford, CA, USA
3: Wu Tsai Neurosciences Institute, Stanford University, Stanford, CA, USA
4: Department of Chemistry, Stanford University, Stanford, CA, USA
†: These authors contributed equally
*Email: congreve@stanford.edu

**Abstract**

Metal halide perovskites possess desirable optical, material, and electrical properties which have had substantial impact on next-generation optoelectronics. However, given the toxicity of lead, alternative lead-free perovskite semiconductors are needed. By fully replacing lead with rare-earth elements, one can simultaneously address toxicity concerns and achieve comparable optoelectronic performance. Here, we demonstrate the synthesis of two-dimensional europium halide perovskite nanoplatelets governed by the formula $L_2EuX_4$ where L is an organic ligand and X is a halide anion. The structure, morphology, and composition of the nanoplatelets are confirmed by XRD, AFM, and XPS. Deep blue-emitting $PEA_2EuBr_4$ perovskite nanoplatelets are synthesized in both the solution- and solid-states with photoluminescence emission centered at 446 nm and CIE color coordinates of (0.1515, 0.0327) and (0.1515, 0.0342), respectively. Additionally, near-ultraviolet $PEA_2EuCl_4$ perovskite nanoplatelets are synthesized in both the solution- and solid-states with photoluminescence emission centered at 400 nm and 401 nm, respectively. Overall, europium halide perovskite nanoplatelets offer a lead-free alternative for deep blue, violet, and near-ultraviolet light emission – charting new pathways for optoelectronics in this energy regime.

**Introduction**

Metal halide perovskite (MHP) semiconductors are promising candidates for next-generation optoelectronics distinguished by their bandgap tunability,[1,2] high charge carrier mobilities,[3,4] sharp color purity,[5,6] high defect tolerance,[7–9] and inexpensive solution processing.[2,10–12] These properties have led to outstanding optoelectronic performance within solar cells,[13–15] lasers,[16,17] and light-emitting diodes (LEDs).[18–21] For instance, perovskite LED performance has rapidly increased in the last decade, achieving external quantum efficiencies exceeding 20%,[19,22–30] luminances surpassing 100,000 cd/m$^2$,[22,31,32] and stabilities in excess of hundreds of hours.[22,27,33,34] In order to achieve these milestones, MHP crystal structures have been carefully engineered to simultaneously optimize both optical and electrical performance.

Bulk MHPs are governed by the formula ABX$_3$, where A is an organic (e.g., methylammonium MA$^+$), inorganic (e.g., Cs$^+$), or hybrid (e.g., Cs$_{0.5}$MA$_{0.5}$$^+$) cation; B is typically Pb$^{2+}$; and X is a halide anion (e.g., I$^-$, Br$^-$, Cl$^-$). While bulk MHPs have enabled significant materials and optoelectronic device performance, deviations from this crystal structure can offer distinct advantages such as stronger exciton binding energy, higher photoluminescence quantum yield (PLQY), and greater environmental stability.[35–37] In particular, two-dimensional (2D) perovskite nanoplatelets have demonstrated excellent promise for emissive MHP materials and optoelectronic devices due to these aforementioned advantages. Generally, 2D perovskite nanoplatelets are formed by substituting bulk A-site cations with a ligand species (e.g., phenethylammonium PEA$^+$) to confine material growth in one dimension.[38–43] For instance, Weidman *et al.* demonstrated that 2D perovskite nanoplatelets based on both Pb$^{2+}$ and Sn$^{2+}$ can achieve tunable absorption and photoluminescence across the visible spectrum.[41] Additionally, our previous work has



demonstrated that 2D lead halide perovskite nanoplatelets can yield efficient violet and ultraviolet-light emitting diodes.[42,43] Yet, most of the reported work on 2D perovskite nanoplatelets has focused on lead-based materials, which raise toxicity concerns.[38–40,42–44] Tin-based 2D perovskite nanoplatelets offer an alternative pathway to address toxicity concerns and have demonstrated comparable optoelectronic performance to lead-based perovskite emitters. For example, Wang *et al.* engineered red tin-based PeLEDs using 2D PEA$_2$SnI$_4$ and TEA$_2$SnI$_4$ (2-thiopheneethyllamine TEA$^+$) emitters which resulted in low turn-on voltages of 2.3 V and sharp color purity denoted by a full width at half maximum (FWHM) of 28 nm.[45] However, their emissions are largely restricted to the red and near-infrared (NIR) regions of the spectrum.[41,45–47] Additionally, blue lead-free light-emitting materials are limited within the broad MHP semiconductor class.[48] Given that the progress of blue perovskite materials and devices has lagged behind that of both lead and lead-free green, red, and NIR perovskite optoelectronics, efficient blue-emitting perovskite emitters that can address toxicity concerns from lead are urgently needed.

Rare-earth halide perovskites offer an alternative material platform that can address toxicity concerns while also achieving luminescence across large regions of the spectrum. For example, Saghy *et al.* reported double perovskite nanocrystals based on rare-earth elements (e.g., Pr$^{3+}$, Ce$^{3+}$, Eu$^{3+}$, Yb$^{3+}$) that display d→f and f→f photoluminescence emission across the ultraviolet (UV), visible, and NIR spectrums.[49] Eu-based rare-earth halide perovskites are particularly promising since these materials have the potential to engineer full-color perovskite-based displays, given that their deep-blue emission is highly aligned with the Rec. 2020 requirement for blue emitters – demarcated by a Commission Internationale de l'Eclairage (CIE) color coordinate of (0.131, 0.046).[50–53] Focusing on Eu-based rare-earth halide perovskites, Huang *et al.* synthesized CsEuCl$_3$



nanocrystals with 435 nm deep-blue emission and achieved PLQYs as high as 5.7% by employing 1-butyl-1-methylpyridinium chloride surface treatment.[50] By employing oleylammonium bromide and trioctylphosphine dibromide ligands, Ha *et al.* achieved 430 nm deep-blue emitting $CsEuBr_3$ nanocrystals with PLQYs as high as 40.5%.[51] However, an exploration of Eu-based 2D perovskite nanoplatelets is missing, which could yield new pathways for lead-free emissive materials and devices in the deep-blue regime and beyond.

In this work, we investigate the synthesis and characterization of both solution- and solid-state 2D europium halide perovskite nanoplatelets. Focusing first on deep-blue emitting nanoplatelets, we develop $PEA_2EuBr_4$ in both the solution- and solid-states, achieving PLQYs as high as 12.8 % and 10.2 %, respectively, while also demonstrating strong alignment with the Rec. 2020 standard for high-definition displays. To understand this class of material, we conduct X-ray diffraction (XRD), atomic force microscopy (AFM), and X-ray photoelectron spectroscopy (XPS) to confirm the structure and composition of the 2D europium halide perovskite nanoplatelets. Time-resolved photoluminescence (TRPL) measurements reveal monoexponential decay lifetimes on the order of 50 ns within our solid-state blue-emitting nanoplatelets. Lastly, by employing chloride instead of bromide within our nanoplatelets, we synthesize both solution- and solid-state $PEA_2EuCl_4$ nanoplatelets which emit at ~400 nm and achieve PLQYs as high as 4.54% and 7.73%, respectively. These chloride-based nanoplatelets offer a lead-free pathway to achieving violet/near-UV emitters within the broad class of MHPs. This work demonstrates the potential of 2D europium halide perovskite nanoplatelets as a platform for next-generation deep-blue, violet, and near-UV emitters while also addressing toxicity concerns within MHP-based materials and devices.



**Results and Discussion**

We utilize europium halide precursors to yield emissive lead-free perovskite nanoplatelets of the form $L_2EuX_4$, where L is an organic ligand and X is a halide anion. Specifically, we synthesize all of our europium halide perovskite nanoplatelets using a nonsolvent crystallization method.[41,54] The synthesis of $PEA_2EuBr_4$ nanoplatelets is shown in **Figure 1A**, where PEABr (phenethylammonium bromide) and $EuBr_2$ are first dissolved at room temperature in N,N-dimethylformamide (DMF) at the same molar concentration. Next, the PEABr and $EuBr_2$ stock solutions are mixed in a 5:1 ratio to form the precursor solution. While theoretically, the stoichiometry of the nanoplatelets calls for a 2:1 ratio of PEABr:$EuBr_2$, we found that the 2:1 ratio of PEABr:EuBr2 dropcasted film does not form solely 2D nanoplatelets (see **Figure S1**). After mixing, 10 μL of the precursor solution is injected into 10 mL of chlorobenzene (CB) under stirring, leading to instantaneous formation of the $PEA_2EuBr_4$ nanoplatelets. The proposed crystal structure of the nanoplatelets formed is shown in **Figure 1B**, where $[EuBr_6]^{4-}$ octahedra are sandwiched between layers of $PEA^+$ cations. In order to confirm this structure, XRD, AFM, and XPS characterizations were performed. To start, the XRD pattern of the dropcasted $PEA_2EuBr_4$ nanoplatelets is shown in **Figure 1C**, as well as the XRD pattern of lead-based $PEA_2PbBr_4$ nanoplatelets[42] for comparison. Both $PEA_2EuBr_4$ and $PEA_2PbBr_4$ show characteristic 2D perovskite peaks, where *(00l)* peaks dominate and appear periodically within the scans.[40–42,55] These *(00l)* peaks are extracted for both $PEA_2EuBr_4$ and $PEA_2PbBr_4$ in Table S1. The *(00l)* peaks of $PEA_2EuBr_4$ are right-shifted as compared to $PEA_2PbBr_4$, implying slight shrinking of the crystal lattice. This is in good agreement with a previous report regarding the smaller ionic radius of $Eu^{2+}$ (117 pm) as compared to that of the larger ionic radius of $Pb^{2+}$ (119 pm).[53] The periodicity of the XRD patterns are 5.5° and 5.6° for $PEA_2PbBr_4$ and $PEA_2EuBr_4$ which equate to stacking distances



of 1.61 nm and 1.58 nm, respectively. Given that the stacking distance is related to a single metal halide octahedra layer and the space the phenethylammonium ligands occupy, the similar stacking distances between $PEA_2EuBr_4$ and $PEA_2PbBr_4$ nanoplatelets is reasonable. Lastly, we perform Pawley refinement to further confirm the crystal structure of $PEA_2EuBr_4$ nanoplatelets as shown in **Figure S2**. Since the *(00l)* peaks dominate the XRD pattern of $PEA_2EuBr_4$ nanoplatelets, the c lattice constant is refined with a value of 1.69 nm. Further, the $PEA_2EuBr_4$ nanoplatelets' stacking distance is not collinear with the c lattice parameter since the $\alpha$ and $\beta$ angles are 99.55 and 105.7 degrees, respectively. Given the c lattice constant with the corresponding $\alpha$ and $\beta$ angles, we calculate the stacking distance as $c \cdot \sin(180-\alpha) \cdot \sin(180-\beta) = 1.60$ nm, which is well-aligned with the 1.58 nm stacking distance derived from the *(00l)* diffraction peak periodicity.



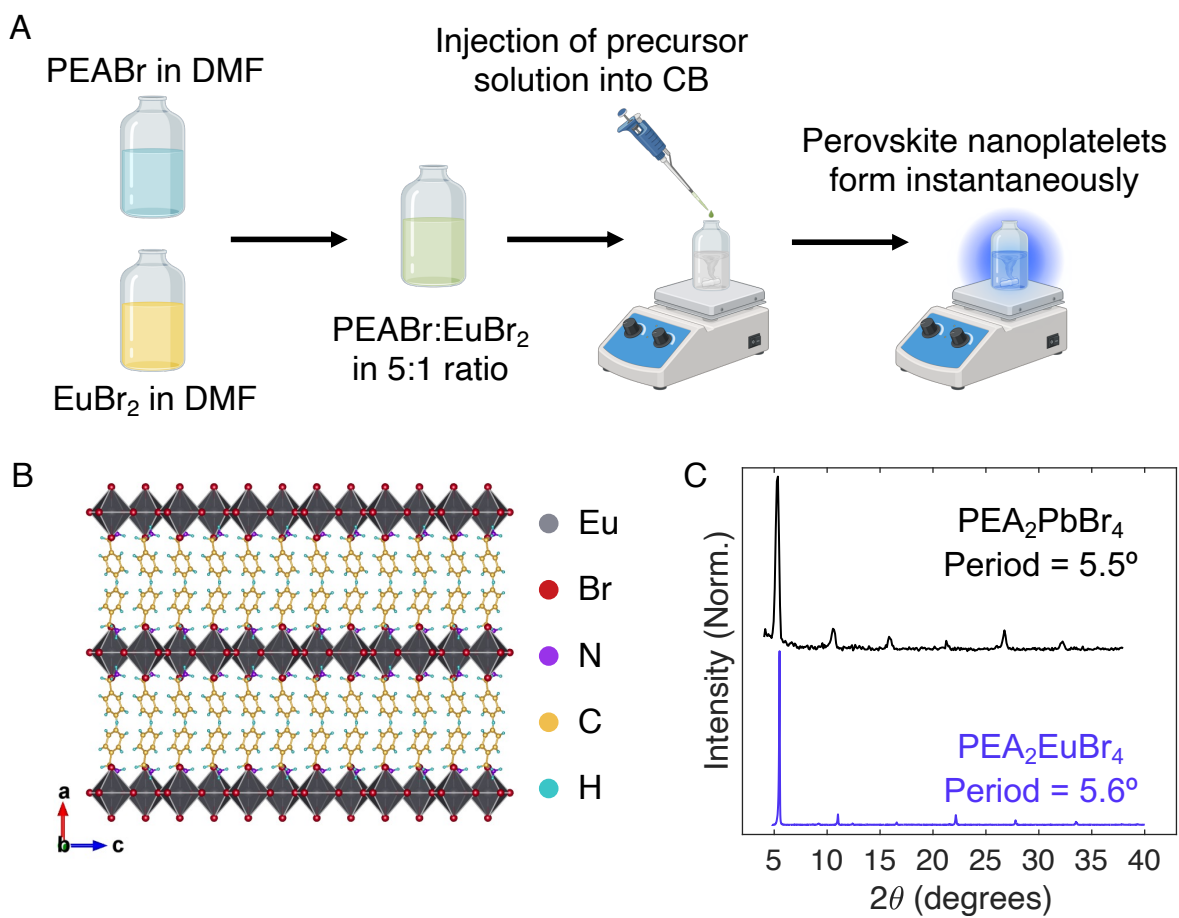

**Figure 1.** Synthesis and structure of PEA$_2$EuBr$_4$ nanoplatelets. **(A)** Schematic of PEA$_2$EuBr$_4$ nanoplatelet synthesis. **(B)** Proposed crystal structure of PEA$_2$EuBr$_4$ nanoplatelets. **(C)** XRD patterns for PEA$_2$PbBr$_4$ and PEA$_2$EuBr$_4$ nanoplatelets. The XRD pattern for PEA$_2$PbBr$_4$ is reproduced from our previous work.[42]

Next, we seek to further characterize the morphology and composition of the PEA$_2$EuBr$_4$ nanoplatelets using AFM and XPS, respectively. **Figure 2A** depicts an AFM image of multiple PEA$_2$EuBr$_4$ nanoplatelets deposited on a Si wafer. To confirm that the smaller background material does not originate from the processing of nanoplatelets, **Figure S3** shows an AFM image of an unprocessed Si wafer (i.e., one with no perovskite nanoplatelets or solutions deposited). The height



profiles of two scans from **Figure 2A** are plotted in **Figure 2B**, which demonstrate nanoplatelet thicknesses of approximately 3 nm across length scales of approximately 2 µm. Given the stacking distance of 1.58 nm from **Figure 1C** and previous reports of single metal halide octahedra layer thicknesses of approximately 0.6 nm,[56,57] the thickness of the phenethylammonium ligand is approximately 1 nm. Therefore, the theoretical thickness of a single $PEA_2EuBr_4$ nanoplatelet is approximated by t = 1 nm + 1 nm + 0.6 nm = 2.6 nm, which is reasonable given the height profiles shown.



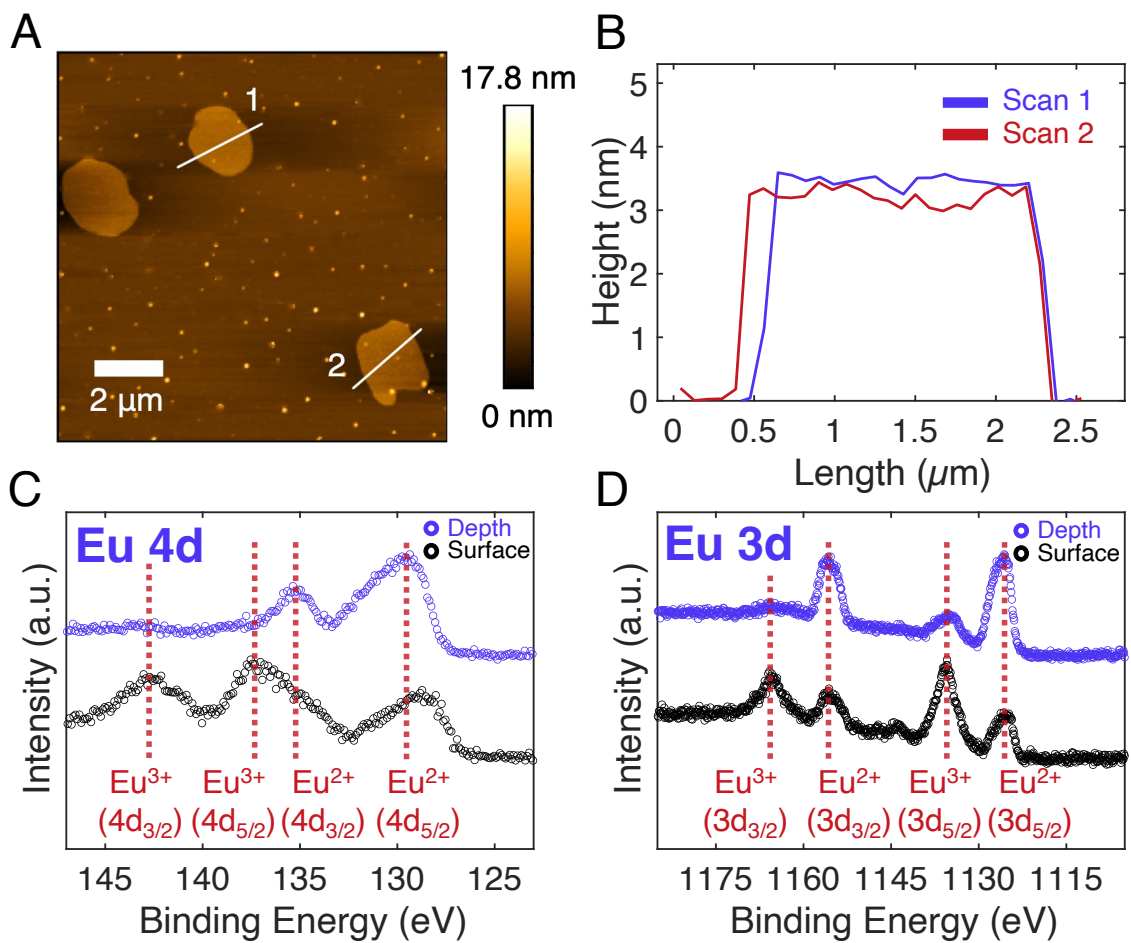

**Figure 2.** Morphology and composition of $PEA_2EuBr_4$ nanoplatelets. **(A)** AFM image of $PEA_2EuBr_4$ nanoplatelets. **(B)** AFM height profiles of $PEA_2EuBr_4$ nanoplatelets. XPS spectra of **(C)** Eu 4d and **(D)** Eu 3d of $PEA_2EuBr_4$ nanoplatelets before and after Ar ion beam treatment.

To confirm the composition of the $PEA_2EuBr_4$ nanoplatelets, we employ XPS and investigate the Eu 4d, Eu 3d, and C 1s spectra. **Figures 2C** and **2D** depict Eu 4d and 3d spectra where the $d_{3/2}$ and $d_{5/2}$ peaks of both $Eu^{3+}$ and $Eu^{2+}$ are detected. Previous investigations comment that the europium in europium halide perovskite materials easily oxidizes from $Eu^{2+}$ to $Eu^{3+}$.[50,51,58] To confirm that the presence of $Eu^{3+}$ can be attributed to surface oxidation, we removed the surface layer with an Ar ion beam and reveal the oxidation state at depth within the dropcasted material. Within both



**Figures 2C** and **2D**, we can see that the $Eu^{3+}/Eu^{2+}$ ratios decrease dramatically after the Ar ion beam process, confirming that $Eu^{3+}$ ions are concentrated on the surface of the dropcasted film and imply that surface oxidation occurred mainly during sample preparation and measurement. Lastly, within **Figure S4**, peaks related to C-C and C-N bonds appear within the C 1s spectra pre- and post-Ar ion beam process, which are a result of the carbon bonds within the phenethylammonium ligands used to synthesize the nanoplatelets.



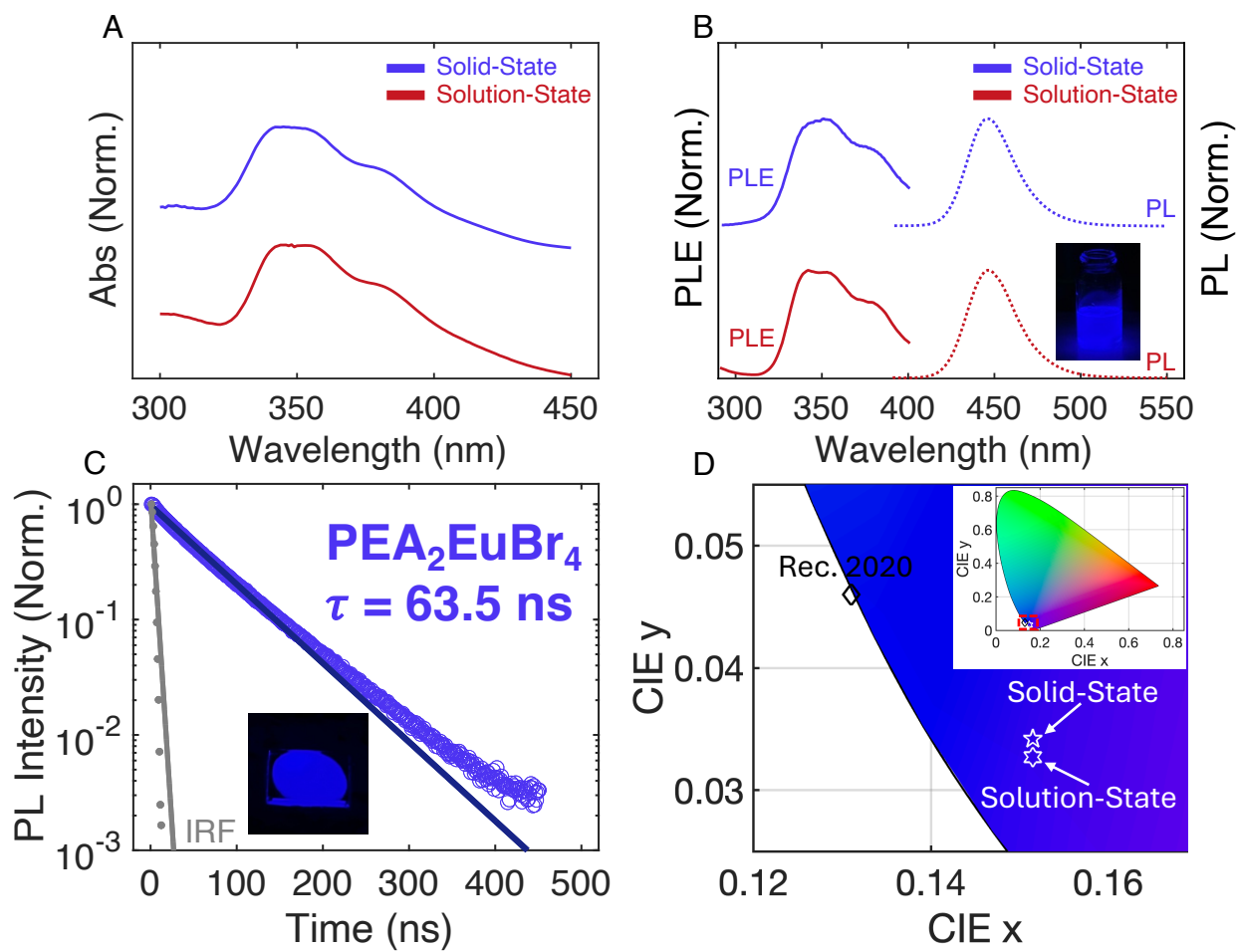

**Figure 3.** Optical properties of PEA$_2$EuBr$_4$ nanoplatelets. **(A)** Absorption of solid- and solution-state PEA$_2$EuBr$_4$ nanoplatelets. **(B)** Photoluminescence excitation (PLE) (left) and photoluminescence (PL) (right) spectra for solid- and solution-state PEA$_2$EuBr$_4$ nanoplatelets. Inset shows solution-state PEA$_2$EuBr$_4$ nanoplatelets under 365 nm light. **(C)** Time-resolved photoluminescence decay curve measured at 446 nm of solid-state PEA$_2$EuBr$_4$ nanoplatelets. $R^2$ = 99.9%. Inset shows solid-state PEA$_2$EuBr$_4$ nanoplatelets under 365 nm light. **(D)** CIE color coordinates for solid- and solution-state PEA$_2$EuBr$_4$ nanoplatelets. Inset shows the entire CIE chromaticity diagram with the PEA$_2$EuBr$_4$ nanoplatelets' coordinates outlined in a red dashed box.



Next, we look toward characterizing the optical properties of the PEA$_2$EuBr$_4$ nanoplatelets. In **Figure 3A**, we show that both solid- and solution-state PEA$_2$EuBr$_4$ nanoplatelets show strong absorption at approximately 350 nm. Further, both solid- and solution-state PEA$_2$EuBr$_4$ nanoplatelets achieve photoluminescence (PL) emission at 446 nm with a FWHM of 33 nm (**Figure 3B**). To confirm the origin of the light emission, we also conduct photoluminescence excitation (PLE) measurements centered at a 446 nm emission wavelength for both solid- and solution-state PEA$_2$EuBr$_4$ nanoplatelets as shown in **Figure 3B**. The absorption and PLE spectrums are well-aligned with respect to one another, which confirm that the light emission originated from the absorption in both solid- and solution-state PEA$_2$EuBr$_4$ nanoplatelets. Given the similarities in optical properties between PEA$_2$EuBr$_4$ and previously reported europium-based MHPs (e.g., CsEuBr$_3$),[50,53] including stoke shifts on the order of 100 nm, discrete shape of both the absorption and PLE spectra, and resulting deep-blue PL, the light emission from PEA$_2$EuBr$_4$ likely originates from the dipole-allowed 4f$^6$5d$^1$→4f$^7$ transition of Eu$^{2+}$. Notably, this mechanism differs from lead- and tin-based perovskite nanoplatelets.[38,41–43] The solution-state 2:1 PEABr:EuBr$_2$ material, from **Figure S5**, shows similar absorption, PL, and PLE spectra of this material to that of the PEA$_2$EuBr$_4$ nanoplatelets in **Figure 3**. This suggests that while the 2:1 PEABr:EuBr$_2$ material does not solely form 2D nanoplatelets, its optical properties are similar to that of PEA$_2$EuBr$_4$ nanoplatelets.

$$PLQY = k_{rad} / (k_{rad} + k_{nonrad}) \tag{1}$$

$$\tau = (k_{rad} + k_{nonrad})^{-1} \tag{2}$$

The PLQYs of solid- and solution-state PEA$_2$EuBr$_4$ nanoplatelets is 10.2% and 12.8%, respectively. To extract the decay lifetime of our solid-state PEA$_2$EuBr$_4$ nanoplatelets, the TRPL decay curve was measured at 446 nm is shown in **Figure 3C,** with the TRPL spectra is shown in **Figure S6**. The decay lifetime is modeled by a monoexponential decay with a lifetime of 63.5 ns, which is similar to that of deep blue-emitting CsEuCl$_3$ nanocrystals (30.9 ns).[50] Next, using equations 1 and 2, we calculate the radiative and nonradiative recombination rates within solid-state PEA$_2$EuBr$_4$ nanoplatelets to be 1.61 x 10$^6$ s$^{-1}$ and 1.41 x 10$^7$ s$^{-1}$, respectively. We summarize these rates, as well as other extracted metrics of PEA$_2$EuBr$_4$, in Table 1. Lastly, **Figure 3D** plots the CIE color coordinates of both solid- and solution-state PEA$_2$EuBr$_4$ nanoplatelets. By comparing the CIE color coordinates for solid- and solution-state PEA$_2$EuBr$_4$ nanoplatelets of (0.1515, 0.0342) and (0.1515, 0.0327), respectively, to the (0.131, 0.046) CIE color coordinate of Rec. 2020's requirement for blue emitters, PEA$_2$EuBr$_4$ nanoplatelets demonstrate strong potential to meet the pressing blue need for full-color perovskite-based displays while simultaneously addressing toxicity concerns from lead-based emitters.

Table 1. Summarized metrics of PEA$_2$EuBr$_4$ nanoplatelets from **Figure 3.**

| PEA$_2$EuBr$_4$ | PL Peak | FWHM | PLQY | $\tau$ | k$_{rad}$ | k$_{nonrad}$ |
|---|---|---|---|---|---|---|
| Solid-State | 446 nm | 33 nm | 10.2 % | 63.5 ns | 1.61 x 10$^6$ s$^{-1}$ | 1.41 x 10$^7$ s$^{-1}$ |
| Solution-State | 446 nm | 33 nm | 12.8 % | N/A | N/A | N/A |

In order to study the ligand's effects on the optical properties of europium bromide nanoplatelets, we first synthesize solution-state nanoplatelets using n-butylammonium bromide (n-BABr) and 2-(4-fluorophenyl)ethylammonium hydrobromide (F-PEABr) ligands. While all of the alternative solution-state nanoplatelets (e.g., (PEA/n-BA)$_2$EuBr$_4$, n-BA$_2$EuBr$_4$, and F-PEA$_2$EuBr$_4$) achieve deep blue PL similar to that of PEA$_2$EuBr$_4$, their PL intensity is comparatively diminished (**Figures S7** and **S8**). However, solution-state (PEA/n-BA)$_2$EuBr$_4$ nanoplatelets show comparable PL intensity to PEA$_2$EuBr$_4$ nanoplatelets and are studied further in both the solution- and solid-state (**Figure S7**). While the PLQYs of solution- and solid-state (PEA/n-BA)$_2$EuBr$_4$ nanoplatelets are lower than that of PEA$_2$EuBr$_4$ nanoplatelets, their CIE coordinates are better aligned with Rec. 2020's blue CIE color coordinate compared to PEA$_2$EuBr$_4$ (**Figure S9**). Extracted metrics of (PEA/n-BA)$_2$EuBr$_4$ nanoplatelets, including monoexponential decay lifetime (**Figure S10**), are summarized in Table S2. Altogether, this suggests that ligand engineering can yield different performance enhancements within europium bromide nanoplatelets and further investigations are warranted. More details regarding these ligand-modified europium bromide nanoplatelets can be found in the **Supporting Information**.

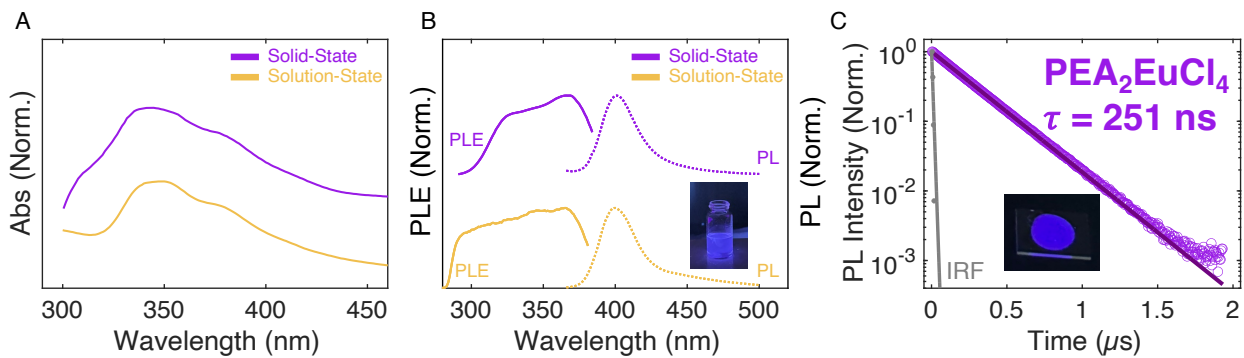

**Figure 4.** Optical properties of PEA$_2$EuCl$_4$ nanoplatelets. **(A)** Absorption of solid- and solution-state PEA$_2$EuCl$_4$ nanoplatelets. **(B)** Photoluminescence excitation (left) and photoluminescence (right) spectra for solid- and solution-state PEA$_2$EuCl$_4$ nanoplatelets. Inset shows solution-state PEA$_2$EuCl$_4$ nanoplatelets under 365 nm light. **(C)** Time-resolved photoluminescence decay curve measured at 401 nm of solid-state PEA$_2$EuCl$_4$ nanoplatelets. $R^2$ = 99.996%. Inset shows solid-state PEA$_2$EuCl$_4$ nanoplatelets under 365 nm light.

Furthermore, we now focus our efforts on achieving beyond blue emitters and push towards the UV spectrum by replacing all of the bromide with chloride within europium halide nanoplatelets to form $PEA_2EuCl_4$ nanoplatelets. The $PEA_2EuCl_4$ nanoplatelet synthesis procedure is identical to that of $PEA_2EuBr_4$ except that the PEACl and $EuCl_2$ precursors are both stirred at 80°C in DMF due to the poor solubility of the chloride precursors. **Figure S11** shows the XRD pattern of the dropcasted $PEA_2EuCl_4$ nanoplatelets. Similar to $PEA_2EuBr_4$ nanoplatelets, signature periodic *(00l)* peaks emerge confirming the formation of 2D perovskite nanoplatelets. The XRD pattern periodicity is 5.5° for $PEA_2EuCl_4$ nanoplatelets which equates to a stacking distance of 1.61 nm. These *(00l)* peaks are extracted for $PEA_2EuCl_4$ and compared to those of $PEA_2EuBr_4$ in Table S3, where the periodic peaks are well-aligned across halide. However, an additional peak emerges at 7.65° for the $PEA_2EuCl_4$ nanoplatelets which could be caused from the poor solubility of the chloride precursors. Additionally, **Figure S12** shows the Pawley refinement of the $PEA_2EuCl_4$ nanoplatelets where the c lattice constant is refined with a value of 1.70 nm. The $\alpha$ and $\beta$ angles are 99.55 and 105.7 degrees, respectively, since the lattice angles were not refined between $PEA_2EuCl_4$ and $PEA_2EuBr_4$ nanoplatelets. Thus, we calculate the stacking distance as c · sin(180-$\alpha$) · sin(180-$\beta$) = 1.61 nm, which agrees with the 1.61 nm stacking distance derived from the *(00l)* diffraction peak periodicity. **Figure S13** shows Eu 4d, Eu 3d, and C 1s spectra of the $PEA_2EuCl_4$ nanoplatelets. Similar to $PEA_2EuBr_4$ nanoplatelets, we can see that both the Eu 4d and 3d spectra in **Figures S13A** and **S13B**, respectively, show the $d_{3/2}$ and $d_{5/2}$ peaks of both $Eu^{3+}$ and $Eu^{2+}$. Finally, **Figure S13C** shows the C 1s spectra which confirm the C-C and C-N bonds within $PEA_2EuCl_4$ nanoplatelets.

Table 2. Summarized metrics of PEA$_2$EuCl$_4$ nanoplatelets from **Figure 4.**

| PEA$_2$EuCl$_4$ | PL Peak | FWHM | PLQY | $\tau$ | $k_{rad}$ | $k_{nonrad}$ |
|---|---|---|---|---|---|---|
| Solid-State | 401 nm | 29 nm | 7.73 % | 251 ns | 3.08 x 10$^5$ s$^{-1}$ | 3.68 x 10$^6$ s$^{-1}$ |
| Solution-State | 400 nm | 29 nm | 4.54 % | N/A | N/A | N/A |

The absorption of solid- and solution-state PEA$_2$EuCl$_4$ nanoplatelets is shown in **Figure 4A**. The PEA$_2$EuCl$_4$ nanoplatelets show strong absorption at approximately 350 nm, similar to that of PEA$_2$EuBr$_4$ nanoplatelets. Both the PL and PLE spectra are shown in **Figure 4B**, where the solid- and solution-state PEA$_2$EuCl$_4$ nanoplatelets achieve PL emission at 401 nm and 400 nm, respectively, with a FWHM of 29 nm. Previous reports note that the 5d orbitals in Eu$^{2+}$ ions are sensitive to the surrounding crystal field.[59–61] Consequently, the replacement of bromide with chloride within these europium halide nanoplatelets could yield a less stabilized 5d orbital in Eu$^{2+}$, increasing the orbital's energy level and resulting in a higher d→f transition energy. This altered d→f transition could explain the blue-shifted emission from PEA$_2$EuCl$_4$ as compared to PEA$_2$EuBr$_4$. The solid- and solution-state PLQYs of PEA$_2$EuCl$_4$ nanoplatelets is 7.73% and 4.54%, respectively. The PLQY of solution-state PEA$_2$EuCl$_4$ nanoplatelets is lower than that of the corresponding solid-state material, which could be attributed to the poor colloidal stability of PEA$_2$EuCl$_4$ due to observed sedimentation between synthesis and PLQY measurements. **Figures 4C** and **S14** show the TRPL decay curve measured at 401 nm and TRPL spectra, respectively, of the solid-state PEA$_2$EuCl$_4$ nanoplatelets, where the extracted monoexponential decay lifetime is 251 ns. Lastly, the radiative and nonradiative recombination rates for PEA$_2$EuCl$_4$ nanoplatelets are 3.08 x 10$^5$ s$^{-1}$ and 3.68 x 10$^6$ s$^{-1}$, respectively, where both recombination rates are suppressed by approximately one order of magnitude as compared to PEA$_2$EuBr$_4$ nanoplatelets. We summarize these rates, as well as other extracted metrics of PEA$_2$EuCl$_4$, in Table 2. All in all, halide engineering within europium halide perovskite nanoplatelets yields new pathways to lead-free near-UV emitters within the broad metal halide perovskite semiconductor class – unlocking new considerations for technologies in this underexplored high energy regime.

**Conclusion**

In this work, we have synthesized europium halide perovskite nanoplatelets of the form $L_2EuX_4$ across various organic ligands and both bromide and chloride anions. Europium bromide nanoplatelets demonstrate deep-blue PL emission with strong alignment with the Rec. 2020 blue emitter requirement and solution- and solid-state PLQYs as high as 12.8% and 10.2%, respectively. Solution- and solid-state europium chloride nanoplatelets exhibit near-UV PL emission centered at 400 nm and 401 nm and comparable PLQYs of 4.54% and 7.73%, respectively. Overall, these results showcase the potential of europium halide perovskite nanoplatelets for deep-blue and near-UV emissive materials and devices and address toxicity concerns found within lead-based emissive materials.

**Methods**

**Materials**

Phenethylammonium bromide (PEABr) and n-butylammonium bromide (n-BABr) were purchased from Greatcell Solar Materials. 2-(4-fluorophenyl)ethylammonium hydrobromide (F-PEABr, >98.0%) was purchased from TCI. Europium (II) bromide ($EuBr_2$, 99.99%) was purchased from Fisher Scientific Company. Europium (II) chloride ($EuCl_2$, 99.99%), anhydrous N,N-Dimethylformamide (DMF, 99.8%), and anhydrous chlorobenzene (CB, 99.8%) were purchased from Sigma-Aldrich. All chemicals were used directly as received.

**Nanoplatelet Synthesis**

The following preparation was conducted inside a nitrogen-filled glovebox. 1.0 M stock solutions of organic ligands (e.g., PEABr) and europium halides (e.g., $EuBr_2$) were prepared by individually dissolving 2.0 mmol of each precursor in 2 mL of DMF and left stirring overnight (at room temperature for europium bromide nanoplatelets and 80°C for europium chloride nanoplatelets). Then, a total of 1 mL of the organic ligand stock solution (e.g., 1 mL of PEABr stock solution or 0.5 mL of both PEABr and n-BABr stock solutions) was combined with 0.2 mL of the europium halide stock solution. 10 μL of this final precursor solution was injected into 10 mL of chlorobenzene undergoing vigorous stirring at room temperature where nanoplatelets were formed immediately.

**Post Synthesis Processing**

Solution-state nanoplatelets were packaged in quartz cuvettes for further characterization. To form solid-state nanoplatelets, the solution-state nanoplatelets were packaged in centrifuge tubes within the nitrogen-filled glovebox and are centrifuged at 8000 rpm for 2 minutes. These same packaged centrifuge tubes were brought back into the glovebox to avoid exposure to ambient. For all characterization except AFM, the supernatant was removed and the pellet was redispersed in approximately 20 μL (for europium bromide nanoplatelets) or 5μL (for europium chloride nanoplatelets) of fresh chlorobenzene. Then, the redispersed pellet was dropcasted onto a glass substrate (20 x 15 mm) and quickly transferred to a 70°C hot plate for 2 hours to remove residual solvents. Samples were transferred to XRD and XPS measurements unpackaged but in an air-free temporary container to minimize exposure to ambient. Samples were encapsulated using a coverslip and ultraviolet epoxy (EPO-TEK OG159-2) before TRPL/PL/PLE/PLQY/Absorption measurements. For AFM measurements, 1 mL of fresh chlorobenzene was added to the supernatant and then 20 μL of the supernatant was dropped on a Si wafer in order to achieve sufficiently low nanoplatelet concentrations and consequently image single nanoplatelets. These samples were also transferred to a 70°C hot plate for 2 hours to remove residual solvents. All substrates were cleaned sequentially via sonication in deionized water, acetone, and isopropanol for 5 min per solvent, dried under compressed air, and transferred to a Jelight UV-Ozone Cleaner to be further treated with $O_2$ plasma for 13 minutes prior to processing.

**Characterization**

XRD patterns were acquired using a PANalytical Empyrean XRD system with Cu Kα radiation at a 45-kV tube voltage and a 40-mA tube current. AFM measurements were gathered in ambient air using a Park Systems FX40 with non-contact mode. XPS data was collected using XPS: PHI VersaProbe 3 with a monochromatized Al(Kα) source and dual-gun neutralizing system. The raw data was processed with MultiPak. The XPS spectra were referenced using the position of the C-C bond at core-level C 1s, corresponding to 284.8 eV. The Ar ion beam treatment was performed for one minute at 2 kV with a current of 1.5 μA. Absorption spectra were collected using an Agilent

Cary 6000i UV/Vis/NIR system with transmission mode. PL and PLE spectra were gathered using a Horiba FluoroLog-3 system with a Xe arc lamp. All PL spectra shown were gathered at a 350 nm excitation wavelength. The PLQY of both the solution- and solid-state europium halide perovskite nanoplatelets were measured in an integrating sphere (Labsphere Inc.) excited by a 375 nm laser (LDH-P-C-375) following de Mello et al.[62] The integrating sphere and spectrometer (Ocean Insight, QE Pro) were calibrated using a radiometric calibration source (Ocean Insight, HL-3P-INT-CAL). Additionally, all components were calibrated against a calibrated Newport photodetector. Time-resolved PL measurements were captured on dropcasted films as a function of time on a Hamamatsu C10627 streak unit coupled with a SP2150i spectrograph and C9300 digital camera. The excitation source was a Hamamatsu pulsed 379 nm laser with a 2 MHz repetition rate (europium bromide nanoplatelets) or 0.5 MHz repetition rate (europium chloride nanoplatelets). The resulting laser fluences for europium bromide nanoplatelets and europium chloride nanoplatelets are 118 nJ/cm$^2$ and 256 nJ/cm$^2$, respectively.

**Acknowledgments**

The authors thank Kyle Frohna and Kieran W.P. Orr for their assistance with the Pawley refinement process and analysis. S.F. acknowledges the support from Stanford University as a Diversifying Academia, Recruiting Excellence (DARE) Fellow, the U.S. Department of Energy (DOE) Building Technologies Office (BTO) as an IBUILD Graduate Research Fellow, Stanford Graduate Fellowship in Science & Engineering (SGF) as a P. Michael Farmwald Fellow, and of the National GEM Consortium as a GEM Fellow. D.M. acknowledges the support from Stanford University as an Enhancing Diversity in Graduate Education (EDGE) Fellow. M.H. acknowledges the support of the Department of Electrical Engineering at Stanford University. H.C. acknowledges the support from a Stanford Interdisciplinary Graduate Fellowship. P.N. acknowledges the support of an SGF as a Gabilan Fellow and the Chevron Fellowship in Energy. T.K.C. acknowledges the support from Stanford University as a Knight-Hennessy Scholar and from the National Science Foundation Graduate Research Fellowship Program. Q.Z. acknowledges the support from the Sunlin & Priscilla Chou Graduate Fellowship and an SGF as a STMicroelectronics Fellow. D.L. acknowledges the support from an SGF as a Gabilan Fellow and from Stanford University as an EDGE Fellow. O.S.L. acknowledges the support from the Swiss National Science Foundation Postdoc.Mobility Fellowship under grant number P500PT 22358. A portion of this work was performed at the Stanford Nano Shared Facilities (SNSF), supported by the National Science Foundation under award ECCS-2026822. This research was performed under an appointment to the Building Technologies Office (BTO) IBUILD Graduate Research Fellowship administered by the Oak Ridge Institute for Science and Education (ORISE) and managed by Oak Ridge National Laboratory (ORNL) for the U.S. Department of Energy (DOE). ORISE is managed by Oak Ridge Associated Universities (ORAU). All opinions expressed in this paper are the author's and do not

**Authors Contributions**

S.F. and D.M. designed and performed experiments, analyzed data, and wrote the manuscript. S.F., D.M., and W.M. synthesized nanoplatelets. D.M., M.H., and Q.Z. performed AFM measurements. H.C. and Q.Z. performed XRD measurements. S.F., D.M., P.N., and T.K.C. performed Absorption, PL, PLE, PLQY, and TRPL characterization measurements. D.M., D.L., and O.S.L. performed XPS measurements. G. H. provided data analysis. D.N.C. supervised the research. All of the authors discussed the results and contributed to the manuscript.

**References**

(1) Unger, E. L.; Kegelmann, L.; Suchan, K.; Sörell, D.; Korte, L.; Albrecht, S. Roadmap and Roadblocks for the Band Gap Tunability of Metal Halide Perovskites. *J. Mater. Chem. A* **2017**, *5* (23), 11401–11409. https://doi.org/10.1039/C7TA00404D.

(2) Zhang, K.; Zhu, N.; Zhang, M.; Wang, L.; Xing, J. Opportunities and Challenges in Perovskite LED Commercialization. *J. Mater. Chem. C* **2021**, *9* (11), 3795–3799. https://doi.org/10.1039/D1TC00232E.

(3) Motta, C.; El-Mellouhi, F.; Sanvito, S. Charge Carrier Mobility in Hybrid Halide Perovskites. *Sci. Rep.* **2015**, *5* (1), 12746. https://doi.org/10.1038/srep12746.

(4) Lim, J.; T. Hörantner, M.; Sakai, N.; M. Ball, J.; Mahesh, S.; K. Noel, N.; Lin, Y.-H.; B. Patel, J.; P. McMeekin, D.; B. Johnston, M.; Wenger, B.; J. Snaith, H. Elucidating the Long-Range Charge Carrier Mobility in Metal Halide Perovskite Thin Films. *Energy Environ. Sci.* **2019**, *12* (1), 169–176. https://doi.org/10.1039/C8EE03395A.

(5) Liang, H.; Yuan, F.; Johnston, A.; Gao, C.; Choubisa, H.; Gao, Y.; Wang, Y.-K.; Sagar, L. K.; Sun, B.; Li, P.; Bappi, G.; Chen, B.; Li, J.; Wang, Y.; Dong, Y.; Ma, D.; Gao, Y.; Liu, Y.; Yuan, M.; Saidaminov, M. I.; Hoogland, S.; Lu, Z.-H.; Sargent, E. H. High Color Purity Lead-Free Perovskite Light-Emitting Diodes via Sn Stabilization. *Adv. Sci.* **2020**, *7* (8), 1903213. https://doi.org/10.1002/advs.201903213.

(6) Liu, A.; Bi, C.; Li, J.; Zhang, M.; Cheng, C.; Binks, D.; Tian, J. High Color-Purity and Efficient Pure-Blue Perovskite Light-Emitting Diodes Based on Strongly Confined Monodispersed Quantum Dots. *Nano Lett.* **2023**, *23* (6), 2405–2411. https://doi.org/10.1021/acs.nanolett.3c00548.

(7) Kang, J.; Wang, L.-W. High Defect Tolerance in Lead Halide Perovskite CsPbBr3. *J. Phys. Chem. Lett.* **2017**, *8* (2), 489–493. https://doi.org/10.1021/acs.jpclett.6b02800.

(8) Kim, G.-W.; Petrozza, A. Defect Tolerance and Intolerance in Metal-Halide Perovskites. *Adv. Energy Mater.* **2020**, *10* (37), 2001959. https://doi.org/10.1002/aenm.202001959.

(9) Steirer, K. X.; Schulz, P.; Teeter, G.; Stevanovic, V.; Yang, M.; Zhu, K.; Berry, J. J. Defect Tolerance in Methylammonium Lead Triiodide Perovskite. *ACS Energy Lett.* **2016**, *1* (2), 360–366. https://doi.org/10.1021/acsenergylett.6b00196.

(10) Huang, F.; Li, M.; Siffalovic, P.; Cao, G.; Tian, J. From Scalable Solution Fabrication of Perovskite Films towards Commercialization of Solar Cells. *Energy Environ. Sci.* **2019**, *12* (2), 518–549. https://doi.org/10.1039/C8EE03025A.

(11) Zhang, J.; Wei, B.; Wang, L.; Yang, X. The Solution-Processed Fabrication of Perovskite Light-Emitting Diodes for Low-Cost and Commercial Applications. *J. Mater. Chem. C* **2021**, *9* (36), 12037–12045. https://doi.org/10.1039/D1TC03385A.

(12) Ahmed, G. H.; Liu, Y.; Bravić, I.; Ng, X.; Heckelmann, I.; Narayanan, P.; Fernández, M. S.; Monserrat, B.; Congreve, D. N.; Feldmann, S. Luminescence Enhancement Due to Symmetry Breaking in Doped Halide Perovskite Nanocrystals. *J. Am. Chem. Soc.* **2022**, *144* (34), 15862–15870. https://doi.org/10.1021/jacs.2c07111.

(13) Green, M. A.; Ho-Baillie, A.; Snaith, H. J. The Emergence of Perovskite Solar Cells. *Nat. Photonics* **2014**, *8* (7), 506–514. https://doi.org/10.1038/nphoton.2014.134.

(14) Frohna, K.; Chosy, C.; Al-Ashouri, A.; Scheler, F.; Chiang, Y.-H.; Dubajic, M.; Parker, J. E.; Walker, J. M.; Zimmermann, L.; Selby, T. A.; Lu, Y.; Roose, B.; Albrecht, S.; Anaya, M.; Stranks, S. D. The Impact of Interfacial Quality and Nanoscale Performance Disorder on the Stability of Alloyed Perovskite Solar Cells. *Nat. Energy* **2025**, *10* (1), 66–76. https://doi.org/10.1038/s41560-024-01660-1.

(15) Liu, C.; Yang, Y.; Chen, H.; Spanopoulos, I.; Bati, A. S. R.; Gilley, I. W.; Chen, J.; Maxwell, A.; Vishal, B.; Reynolds, R. P.; Wiggins, T. E.; Wang, Z.; Huang, C.; Fletcher, J.; Liu, Y.; Chen, L. X.; De Wolf, S.; Chen, B.; Zheng, D.; Marks, T. J.; Facchetti, A.; Sargent, E. H.; Kanatzidis, M. G. Two-Dimensional Perovskitoids Enhance Stability in Perovskite Solar Cells. *Nature* **2024**, *633* (8029), 359–364. https://doi.org/10.1038/s41586-024-07764-8.

(16) Zhang, Q.; Shang, Q.; Su, R.; Do, T. T. H.; Xiong, Q. Halide Perovskite Semiconductor Lasers: Materials, Cavity Design, and Low Threshold. *Nano Lett.* **2021**, *21* (5), 1903–1914. https://doi.org/10.1021/acs.nanolett.0c03593.

(17) Zou, C.; Ren, Z.; Hui, K.; Wang, Z.; Fan, Y.; Yang, Y.; Yuan, B.; Zhao, B.; Di, D. Electrically Driven Lasing from a Dual-Cavity Perovskite Device. *Nature* **2025**, *645* (8080), 369–374. https://doi.org/10.1038/s41586-025-09457-2.

(18) Fakharuddin, A.; Gangishetty, M. K.; Abdi-Jalebi, M.; Chin, S.-H.; bin Mohd Yusoff, A. R.; Congreve, D. N.; Tress, W.; Deschler, F.; Vasilopoulou, M.; Bolink, H. J. Perovskite Light-Emitting Diodes. *Nat. Electron.* **2022**, *5* (4), 203–216. https://doi.org/10.1038/s41928-022-00745-7.

(19) Ma, D.; Lin, K.; Dong, Y.; Choubisa, H.; Proppe, A. H.; Wu, D.; Wang, Y.-K.; Chen, B.; Li, P.; Fan, J. Z.; Yuan, F.; Johnston, A.; Liu, Y.; Kang, Y.; Lu, Z.-H.; Wei, Z.; Sargent, E. H. Distribution Control Enables Efficient Reduced-Dimensional Perovskite LEDs. *Nature* **2021**, *599* (7886), 594–598. https://doi.org/10.1038/s41586-021-03997-z.

(20) Chen, Y.; Wang, R.; Kusch, G.; Xu, B.; Hao, C.; Xue, C.; Cheng, L.; Zhu, L.; Wang, J.; Li, H.; Oliver, R. A.; Wang, N.; Huang, W.; Wang, J. All-Site Alloyed Perovskite for Efficient and Bright Blue Light-Emitting Diodes. *Nat. Commun.* **2025**, *16* (1), 3254. https://doi.org/10.1038/s41467-025-58470-6.

(21) Dong, J.; Zhao, B.; Ji, H.; Zang, Z.; Kong, L.; Chu, C.; Han, D.; Wang, J.; Fu, Y.; Zhang, Z.-H.; Yang, Y.; Zhang, L.; Yang, X.; Wang, N. Multivalent-Effect Immobilization of Reduced-Dimensional Perovskites for Efficient and Spectrally Stable Deep-Blue Light-Emitting Diodes. *Nat. Nanotechnol.* **2025**, *20* (4), 507–514. https://doi.org/10.1038/s41565-024-01852-6.

(22) Kim, J. S.; Heo, J.-M.; Park, G.-S.; Woo, S.-J.; Cho, C.; Yun, H. J.; Kim, D.-H.; Park, J.; Lee, S.-C.; Park, S.-H.; Yoon, E.; Greenham, N. C.; Lee, T.-W. Ultra-Bright, Efficient and Stable Perovskite Light-Emitting Diodes. *Nature* **2022**, *611* (7937), 688–694. https://doi.org/10.1038/s41586-022-05304-w.

(23) Xie, M.; Guo, J.; Zhang, X.; Bi, C.; Zhang, L.; Chu, Z.; Zheng, W.; You, J.; Tian, J. High-Efficiency Pure-Red Perovskite Quantum-Dot Light-Emitting Diodes. *Nano Lett.* **2022**, *22* (20), 8266–8273. https://doi.org/10.1021/acs.nanolett.2c03062.

(24) Liu, Y.; Tao, C.; Cao, Y.; Chen, L.; Wang, S.; Li, P.; Wang, C.; Liu, C.; Ye, F.; Hu, S.; Xiao, M.; Gao, Z.; Gui, P.; Yao, F.; Dong, K.; Li, J.; Hu, X.; Cong, H.; Jia, S.; Wang, T.; Wang, J.; Li, G.; Huang, W.; Ke, W.; Wang, J.; Fang, G. Synergistic Passivation and Stepped-Dimensional Perovskite Analogs Enable High-Efficiency near-Infrared Light-Emitting Diodes. *Nat. Commun.* **2022**, *13* (1), 7425. https://doi.org/10.1038/s41467-022-35218-0.

(25) Jiang, J.; Chu, Z.; Yin, Z.; Li, J.; Yang, Y.; Chen, J.; Wu, J.; You, J.; Zhang, X. Red Perovskite Light-Emitting Diodes with Efficiency Exceeding 25% Realized by Co-Spacer Cations. *Adv. Mater.* **2022**, *34* (36), 2204460. https://doi.org/10.1002/adma.202204460.

(26) Lin, K.; Xing, J.; Quan, L. N.; de Arquer, F. P. G.; Gong, X.; Lu, J.; Xie, L.; Zhao, W.; Zhang, D.; Yan, C.; Li, W.; Liu, X.; Lu, Y.; Kirman, J.; Sargent, E. H.; Xiong, Q.; Wei, Z.

Perovskite Light-Emitting Diodes with External Quantum Efficiency Exceeding 20 per Cent. *Nature* **2018**, *562* (7726), 245–248. https://doi.org/10.1038/s41586-018-0575-3.

(27) Guo, B.; Lai, R.; Jiang, S.; Zhou, L.; Ren, Z.; Lian, Y.; Li, P.; Cao, X.; Xing, S.; Wang, Y.; Li, W.; Zou, C.; Chen, M.; Hong, Z.; Li, C.; Zhao, B.; Di, D. Ultrastable Near-Infrared Perovskite Light-Emitting Diodes. *Nat. Photonics* **2022**, *16* (9), 637–643. https://doi.org/10.1038/s41566-022-01046-3.

(28) Feng, S.-C.; Shen, Y.; Hu, X.-M.; Su, Z.-H.; Zhang, K.; Wang, B.-F.; Cao, L.-X.; Xie, F.-M.; Li, H.-Z.; Gao, X.; Tang, J.-X.; Li, Y.-Q. Efficient and Stable Red Perovskite Light-Emitting Diodes via Thermodynamic Crystallization Control. *Adv. Mater.* **2024**, *36* (44), 2410255. https://doi.org/10.1002/adma.202410255.

(29) Sun, S.-Q.; Tai, J.-W.; He, W.; Yu, Y.-J.; Feng, Z.-Q.; Sun, Q.; Tong, K.-N.; Shi, K.; Liu, B.-C.; Zhu, M.; Wei, G.; Fan, J.; Xie, Y.-M.; Liao, L.-S.; Fung, M.-K. Enhancing Light Outcoupling Efficiency via Anisotropic Low Refractive Index Electron Transporting Materials for Efficient Perovskite Light-Emitting Diodes. *Adv. Mater.* **2024**, *36* (24), 2400421. https://doi.org/10.1002/adma.202400421.

(30) Li, M.; Yang, Y.; Kuang, Z.; Hao, C.; Wang, S.; Lu, F.; Liu, Z.; Liu, J.; Zeng, L.; Cai, Y.; Mao, Y.; Guo, J.; Tian, H.; Xing, G.; Cao, Y.; Ma, C.; Wang, N.; Peng, Q.; Zhu, L.; Huang, W.; Wang, J. Acceleration of Radiative Recombination for Efficient Perovskite LEDs. *Nature* **2024**, *630* (8017), 631–635. https://doi.org/10.1038/s41586-024-07460-7.

(31) Shen, G.; Zhang, Y.; Juarez, J.; Contreras, H.; Sindt, C.; Xu, Y.; Kline, J.; Barlow, S.; Reichmanis, E.; Marder, S. R.; Ginger, D. S. Increased Brightness and Reduced Efficiency Droop in Perovskite Quantum Dot Light-Emitting Diodes Using Carbazole-Based Phosphonic Acid Interface Modifiers. *ACS Nano* **2025**. https://doi.org/10.1021/acsnano.4c13036.

(32) Fernández, S.; Michaels, W.; Hu, M.; Narayanan, P.; Murrietta, N.; Gallegos, A. O.; Ahmed, G. H.; Lyu, J.; Gangishetty, M. K.; Congreve, D. N. Trade-off between Efficiency and Stability in Mn2+-Doped Perovskite Light-Emitting Diodes. *Device* **2023**, *1* (2). https://doi.org/10.1016/j.device.2023.100017.

(33) Li, H.; Lin, H.; Ouyang, D.; Yao, C.; Li, C.; Sun, J.; Song, Y.; Wang, Y.; Yan, Y.; Wang, Y.; Dong, Q.; Choy, W. C. H. Efficient and Stable Red Perovskite Light-Emitting Diodes with Operational Stability >300 h. *Adv. Mater.* **2021**, *33* (15), 2008820. https://doi.org/10.1002/adma.202008820.

(34) Han, B.; Yuan, S.; Cai, B.; Song, J.; Liu, W.; Zhang, F.; Fang, T.; Wei, C.; Zeng, H. Green Perovskite Light-Emitting Diodes with 200 Hours Stability and 16% Efficiency: Cross-Linking Strategy and Mechanism. *Adv. Funct. Mater.* **2021**, *31* (26), 2011003. https://doi.org/10.1002/adfm.202011003.

(35) Biswas, A.; Bakthavatsalam, R.; Kundu, J. Efficient Exciton to Dopant Energy Transfer in Mn2+-Doped (C4H9NH3)2PbBr4 Two-Dimensional (2D) Layered Perovskites. *Chem. Mater.* **2017**, *29* (18), 7816–7825. https://doi.org/10.1021/acs.chemmater.7b02429.

(36) Wang, Q.; Liu, X.-D.; Qiu, Y.-H.; Chen, K.; Zhou, L.; Wang, Q.-Q. Quantum Confinement Effect and Exciton Binding Energy of Layered Perovskite Nanoplatelets. *AIP Adv.* **2018**, *8* (2). https://doi.org/10.1063/1.5020836.

(37) Otero-Martínez, C.; Ye, J.; Sung, J.; Pastoriza-Santos, I.; Pérez-Juste, J.; Xia, Z.; Rao, A.; Hoye, R. L. Z.; Polavarapu, L. Colloidal Metal-Halide Perovskite Nanoplatelets: Thickness-Controlled Synthesis, Properties, and Application in Light-Emitting Diodes. *Adv. Mater.* **2022**, *34* (10), 2107105. https://doi.org/10.1002/adma.202107105.

(38) Yang, S.; Niu, W.; Wang, A.-L.; Fan, Z.; Chen, B.; Tan, C.; Lu, Q.; Zhang, H. Ultrathin Two-Dimensional Organic–Inorganic Hybrid Perovskite Nanosheets with Bright, Tunable Photoluminescence and High Stability. *Angew. Chem. Int. Ed.* **2017**, *56* (15), 4252–4255. https://doi.org/10.1002/anie.201701134.
(39) Ma, P.; Zhuang, Y.; Lin, H.; You, G.; Xu, G.; Cai, B. Quantum Yield and Stability Improvement of Two-Dimensional Perovskite Via a Second Fluorinated Insulator Layer. *Adv. Mater. Interfaces* **2023**, *10* (9), 2202003. https://doi.org/10.1002/admi.202202003.
(40) Liang, D.; Peng, Y.; Fu, Y.; Shearer, M. J.; Zhang, J.; Zhai, J.; Zhang, Y.; Hamers, R. J.; Andrew, T. L.; Jin, S. Color-Pure Violet-Light-Emitting Diodes Based on Layered Lead Halide Perovskite Nanoplates. *ACS Nano* **2016**, *10* (7), 6897–6904. https://doi.org/10.1021/acsnano.6b02683.
(41) Weidman, M. C.; Seitz, M.; Stranks, S. D.; Tisdale, W. A. Highly Tunable Colloidal Perovskite Nanoplatelets through Variable Cation, Metal, and Halide Composition. *ACS Nano* **2016**, *10* (8), 7830–7839. https://doi.org/10.1021/acsnano.6b03496.
(42) Hu, M.; Fernández, S.; Zhou, Q.; Narayanan, P.; Saini, B.; Schloemer, T. H.; Lyu, J.; Gallegos, A. O.; Ahmed, G. H.; Congreve, D. N. Water Additives Improve the Efficiency of Violet Perovskite Light-Emitting Diodes. *Matter* **2023**, *6* (7), 2356–2367. https://doi.org/10.1016/j.matt.2023.05.018.
(43) Hu, M.; Lyu, J.; Murrietta, N.; Fernández, S.; Michaels, W.; Zhou, Q.; Narayanan, P.; Congreve, D. N. 2D Mixed Halide Perovskites for Ultraviolet Light-Emitting Diodes. *Device* **2024**, 100511. https://doi.org/10.1016/j.device.2024.100511.
(44) Hong, K.; Le, Q. V.; Kim, S. Y.; Jang, H. W. Low-Dimensional Halide Perovskites: Review and Issues. *J. Mater. Chem. C* **2018**, *6* (9), 2189–2209. https://doi.org/10.1039/C7TC05658C.
(45) Wang, Z.; Wang, F.; Zhao, B.; Qu, S.; Hayat, T.; Alsaedi, A.; Sui, L.; Yuan, K.; Zhang, J.; Wei, Z.; Tan, Z. Efficient Two-Dimensional Tin Halide Perovskite Light-Emitting Diodes via a Spacer Cation Substitution Strategy. *J. Phys. Chem. Lett.* **2020**, *11* (3), 1120–1127. https://doi.org/10.1021/acs.jpclett.9b03565.
(46) Han, D.; Wang, J.; Agosta, L.; Zang, Z.; Zhao, B.; Kong, L.; Lu, H.; Mosquera-Lois, I.; Carnevali, V.; Dong, J.; Zhou, J.; Ji, H.; Pfeifer, L.; Zakeeruddin, S. M.; Yang, Y.; Wu, B.; Rothlisberger, U.; Yang, X.; Grätzel, M.; Wang, N. Tautomeric Mixture Coordination Enables Efficient Lead-Free Perovskite LEDs. *Nature* **2023**, *622* (7983), 493–498. https://doi.org/10.1038/s41586-023-06514-6.
(47) Huang, Y.; Li, D.; Guo, J.; Miao, Z.; Li, S.; Song, Z.; Yin, W.; Zhang, X. Ligand and Stoichiometry Engineering Unlock Narrowband Red Emission in Colloidal PEA2SnI4 Nanoplatelets. *Inorg. Chem.* **2025**, *64* (26), 13463–13469. https://doi.org/10.1021/acs.inorgchem.5c02036.
(48) Ko, P. K.; Ge, J.; Ding, P.; Chen, D.; Tsang, H. L. T.; Kumar, N.; Halpert, J. E. The Deepest Blue: Major Advances and Challenges in Deep Blue Emitting Quasi-2D and Nanocrystalline Perovskite LEDs. *Adv. Mater.* **2025**, *37* (23), 2407764. https://doi.org/10.1002/adma.202407764.
(49) Saghy, P.; Brown, A. M.; Chu, C.; Dube, L. C.; Zheng, W.; Robinson, J. R.; Chen, O. Lanthanide Double Perovskite Nanocrystals with Emissions Covering the UV-C to NIR Spectral Range. *Adv. Opt. Mater.* **2023**, *11* (12), 2300277. https://doi.org/10.1002/adom.202300277.

(50) Huang, J.; Lei, T.; Siron, M.; Zhang, Y.; Yu, S.; Seeler, F.; Dehestani, A.; Quan, L. N.; Schierle-Arndt, K.; Yang, P. Lead-Free Cesium Europium Halide Perovskite Nanocrystals. *Nano Lett.* **2020**, *20* (5), 3734–3739. https://doi.org/10.1021/acs.nanolett.0c00692.

(51) Ha, J.; Yeon, S.; Lee, J.; Lee, H.; Cho, H. Revealing the Role of Organic Ligands in Deep-Blue-Emitting Colloidal Europium Bromide Perovskite Nanocrystals. *ACS Nano* **2024**, *18* (46), 31891–31902. https://doi.org/10.1021/acsnano.4c09018.

(52) Han, T.-H.; Jang, K. Y.; Dong, Y.; Friend, R. H.; Sargent, E. H.; Lee, T.-W. A Roadmap for the Commercialization of Perovskite Light Emitters. *Nat. Rev. Mater.* **2022**, *7* (10), 757–777. https://doi.org/10.1038/s41578-022-00459-4.

(53) Luo, J.; Yang, L.; Tan, Z.; Xie, W.; Sun, Q.; Li, J.; Du, P.; Xiao, Q.; Wang, L.; Zhao, X.; Niu, G.; Gao, L.; Jin, S.; Tang, J. Efficient Blue Light Emitting Diodes Based On Europium Halide Perovskites. *Adv. Mater.* **2021**, *33* (38), 2101903. https://doi.org/10.1002/adma.202101903.

(54) Schmidt, L. C.; Pertegás, A.; González-Carrero, S.; Malinkiewicz, O.; Agouram, S.; Mínguez Espallargas, G.; Bolink, H. J.; Galian, R. E.; Pérez-Prieto, J. Nontemplate Synthesis of CH3NH3PbBr3 Perovskite Nanoparticles. *J. Am. Chem. Soc.* **2014**, *136* (3), 850–853. https://doi.org/10.1021/ja4109209.

(55) Kitazawa, N.; Watanabe, Y. Optical Properties of Natural Quantum-Well Compounds (C6H5-CnH2n-NH3)2PbBr4 (*n*=1–4). *J. Phys. Chem. Solids* **2010**, *71* (5), 797–802. https://doi.org/10.1016/j.jpcs.2010.02.006.

(56) Tyagi, P.; Arveson, S. M.; Tisdale, W. A. Colloidal Organohalide Perovskite Nanoplatelets Exhibiting Quantum Confinement. *J. Phys. Chem. Lett.* **2015**, *6* (10), 1911–1916. https://doi.org/10.1021/acs.jpclett.5b00664.

(57) Pathak, S.; Sakai, N.; Wisnivesky Rocca Rivarola, F.; Stranks, S. D.; Liu, J.; Eperon, G. E.; Ducati, C.; Wojciechowski, K.; Griffiths, J. T.; Haghighirad, A. A.; Pellaroque, A.; Friend, R. H.; Snaith, H. J. Perovskite Crystals for Tunable White Light Emission. *Chem. Mater.* **2015**, *27* (23), 8066–8075. https://doi.org/10.1021/acs.chemmater.5b03769.

(58) Jin, M.; Zeng, Z.; Fu, H.; Wang, S.; Yin, Z.; Zhai, X.; Zhang, Q.; Du, Y. Strain-Negligible Eu2+ Doping Enabled Color-Tunable Harsh Condition-Resistant Perovskite Nanocrystals for Superior Light-Emitting Diodes. *JACS Au* **2023**, *3* (1), 216–226. https://doi.org/10.1021/jacsau.2c00593.

(59) Woo, H. Y.; Yu, M. Y.; Kim, S. H.; Lee, D. W.; Choi, Y.; Kim, Y.; Park, G.; Choi, H.; Paik, T. Divalent Europium-Containing Colloidal Metal Halide Nanocrystals for Light-Emitting Applications. *Nano Converg.* **2025**, *12* (1), 31. https://doi.org/10.1186/s40580-025-00496-z.

(60) Zheng, T.; Sójka, M.; Woźny, P.; Martín, I. R.; Lavín, V.; Zych, E.; Lis, S.; Du, P.; Luo, L.; Runowski, M. Supersensitive Ratiometric Thermometry and Manometry Based on Dual-Emitting Centers in Eu2+/Sm2+-Doped Strontium Tetraborate Phosphors. *Adv. Opt. Mater.* **2022**, *10* (20), 2201055. https://doi.org/10.1002/adom.202201055.

(61) Qin, X.; Liu, X.; Huang, W.; Bettinelli, M.; Liu, X. Lanthanide-Activated Phosphors Based on 4f-5d Optical Transitions: Theoretical and Experimental Aspects. *Chem. Rev.* **2017**, *117* (5), 4488–4527. https://doi.org/10.1021/acs.chemrev.6b00691.

(62) de Mello, J. C.; Wittmann, H. F.; Friend, R. H. An Improved Experimental Determination of External Photoluminescence Quantum Efficiency. *Adv. Mater.* **1997**, *9* (3), 230–232. https://doi.org/10.1002/adma.19970090308.

# Supporting Information

# Lead-Free Europium Halide Perovskite Nanoplatelets

*Sebastian Fernández,[1,†] Divine Mbachu,[1,†] Manchen Hu,[1] Han Cui,[2,3] William Michaels,[1] Pournima Narayanan,[1,4] Tyler K. Colenbrander,[1] Qi Zhou,[1] Da Lin,[1,2] Ona Segura Lecina,[1] Guosong Hong,[2,3] Daniel N. Congreve*,[1]*

1: Department of Electrical Engineering, Stanford University, Stanford, CA, USA
2: Department of Materials Science and Engineering, Stanford University, Stanford, CA, USA
3: Wu Tsai Neurosciences Institute, Stanford University, Stanford, CA, USA
4: Department of Chemistry, Stanford University, Stanford, CA, USA
†: These authors contributed equally
*Email: congreve@stanford.edu

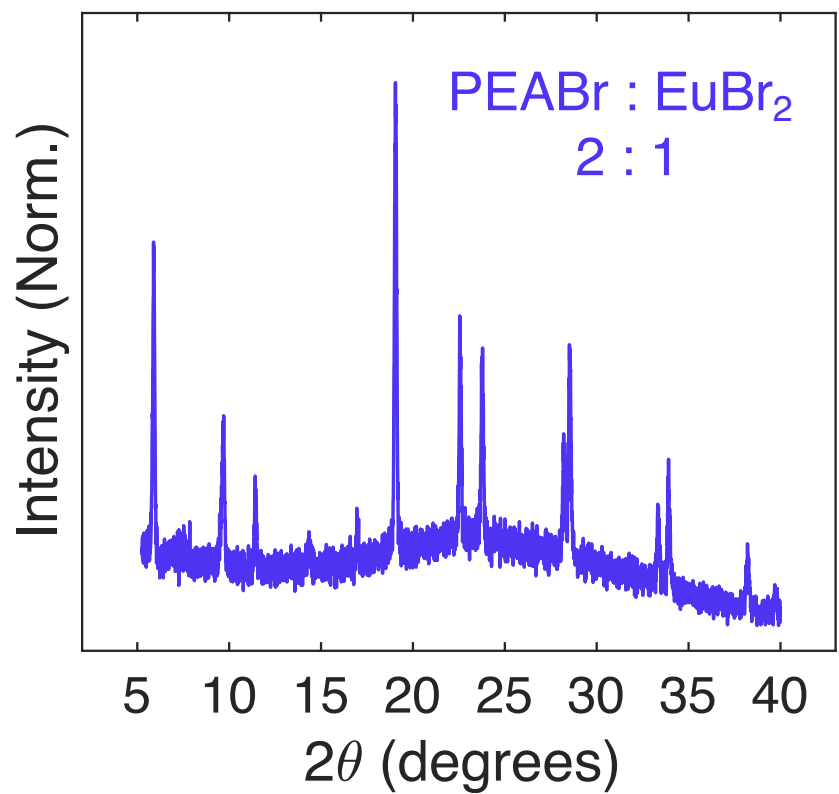

**Figure S1.** XRD pattern of 2:1 PEABr:EuBr$_2$ dropcasted film. Comparing this XRD pattern with that of PEA$_2$EuBr$_4$ in **Figure 1C**, this pattern is not dominated by periodic *(00l)* peaks, which are characteristic of 2D perovskites.

Table S1. Periodic XRD peak positions of both PEA$_2$EuBr$_4$ and PEA$_2$PbBr$_4$ nanoplatelets.

| Perovskite | Peak 1 | Peak 2 | Peak 3 | Peak 4 | Peak 5 | Peak 6 |
|---|---|---|---|---|---|---|
| PEA$_2$EuBr$_4$ | 5.51 º | 11.0 º | 16.6 º | 22.2 º | 27.8 º | 33.5 º |
| PEA$_2$PbBr$_4$ | 5.35 º | 10.7 º | 15.9 º | 21.3 º | 26.8 º | 32.3 º |

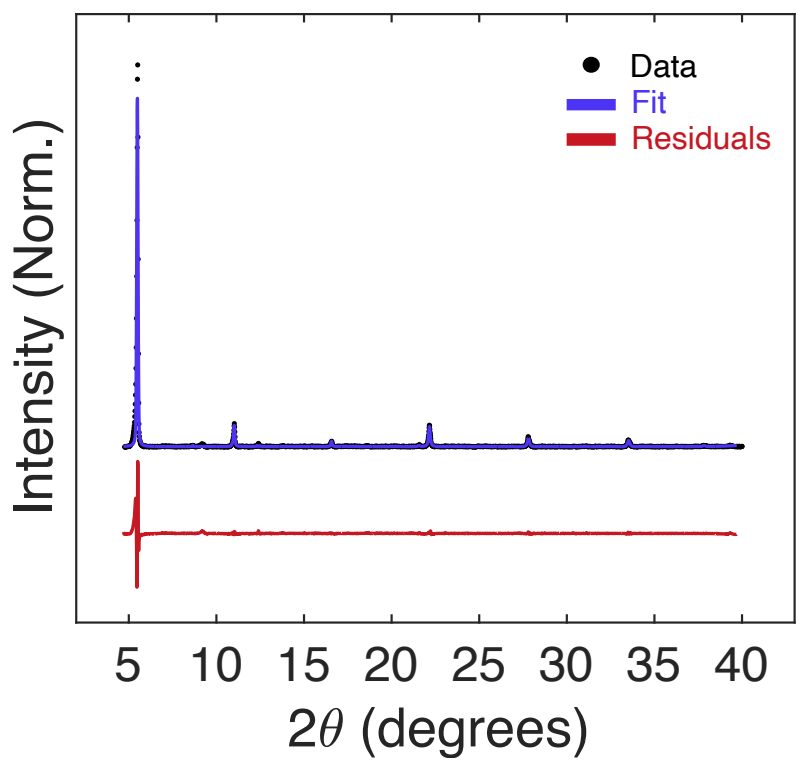

**Figure S2.** Pawley refinement of the XRD pattern for PEA$_2$EuBr$_4$ nanoplatelets.

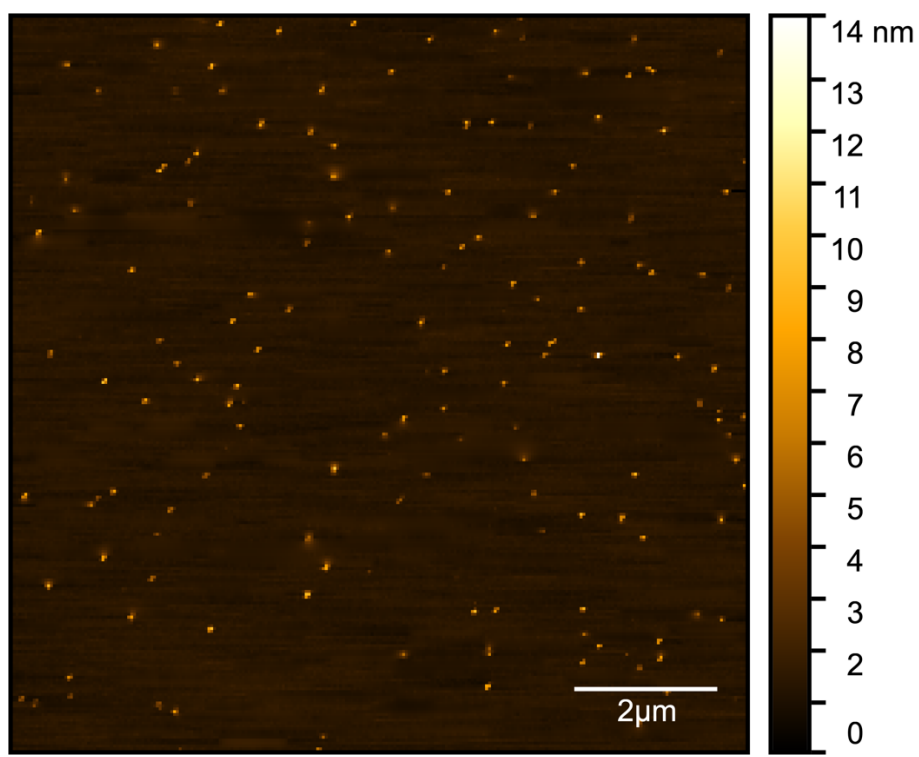

**Figure S3.** AFM image of Si wafer with no perovskite nanoplatelets deposited.

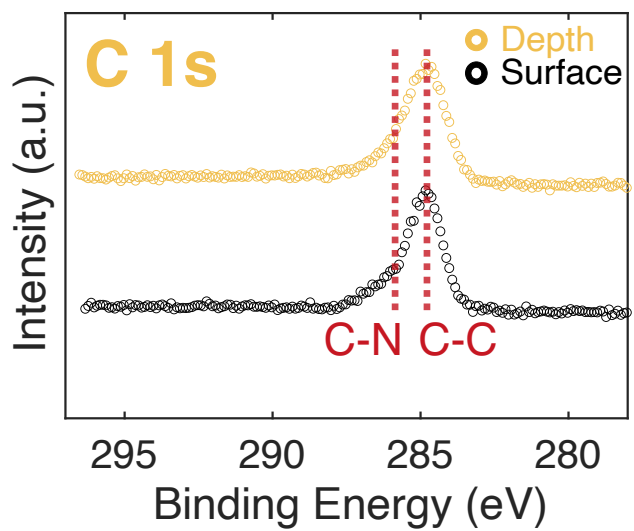

**Figure S4.** XPS spectra of C 1s of PEA$_2$EuBr$_4$ nanoplatelets before and after Ar ion beam treatment.

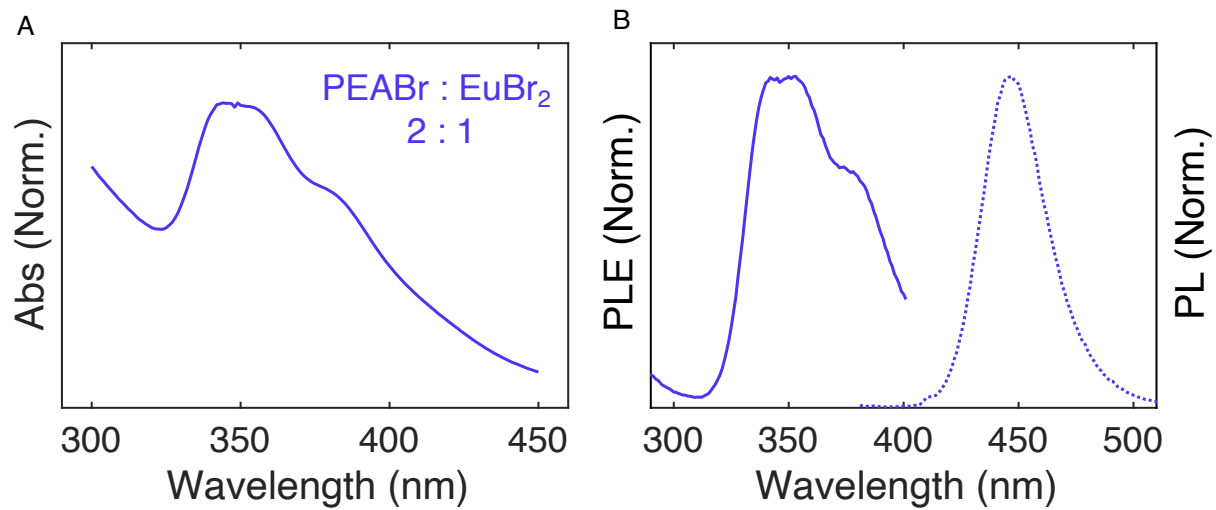

**Figure S5.** Absorption **(A)**, photoluminescence excitation and photoluminescence **(B)** spectra of solution-state 2:1 PEABr:EuBr$_2$.

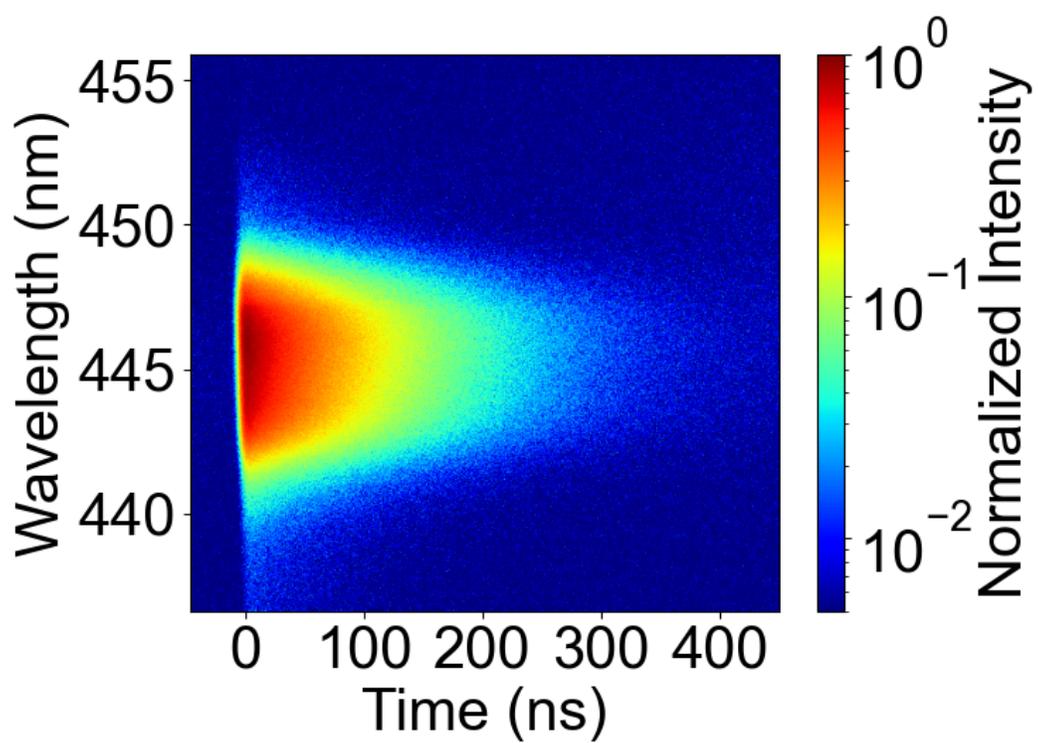

**Figure S6.** Time-resolved photoluminescence spectra of solid-state PEA$_2$EuBr$_4$ nanoplatelets. Note: a 445 nm bandpass filter was used within this optical set-up.

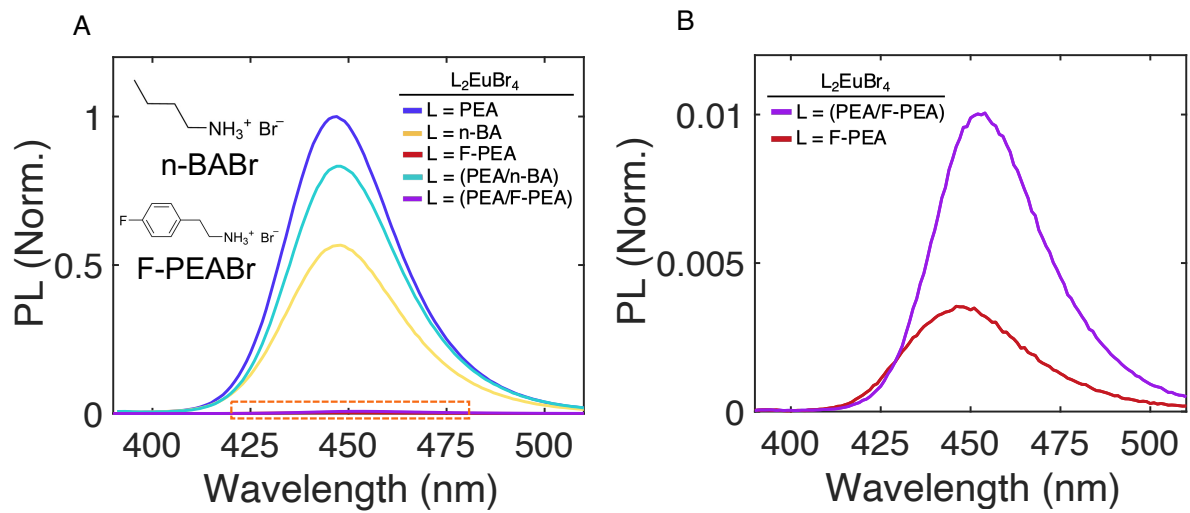

**Figure S7.** Photoluminescence spectra of solution-state $L_2EuBr_4$ across various organic ligands. **(A)** Photoluminescence spectra of $L_2EuBr_4$ normalized to that of $PEA_2EuBr_4$. Photoluminescence intensity of $L_2EuBr_4$ are also normalized to their respective absorption at the 350 nm excitation wavelength. Inset shows the structures of n-BABr and F-PEABr. Dashed red rectangle denotes the reduced photoluminescence intensity from $(PEA/F-PEA)_2EuBr_4$ and $F-PEA_2EuBr_4$. **(B)** Zoomed in photoluminescence spectra of $(PEA/F-PEA)_2EuBr_4$ and $F-PEA_2EuBr_4$. Y-axis scale denotes significant reduction in intensity as compared to other $L_2EuBr_4$ nanoplatelets.

In an effort to explore the formation of alternative europium bromide nanoplatelets, we also employ n-butylammonium bromide (n-BABr) and 2-(4-fluorophenyl)ethylammonium hydrobromide (F-PEABr) ligands to synthesize solution-state nanoplatelets of the formula $L_2EuBr_4$. We first compare the relative PL intensities across $L_2EuBr_4$ nanoplatelets (where L = n-BA, F-PEA, and equimolar ratios of n-BA & PEA and F-PEA & PEA) to $PEA_2EuBr_4$ in **Figure S7**. From **Figure S7A**, we normalize all PL spectra to that of $PEA_2EuBr_4$ and note that the PL intensity of $PEA_2EuBr_4$ is stronger than that of all other synthesized europium halide nanoplatelets. However, the solution-state $(PEA/n-BA)_2EuBr_4$ nanoplatelets (equimolar mixture of phenethylammonium and n-butylammonium ligands) show comparable PL intensity at 448 nm and will be studied further. Additionally, **Figure S7B** shows that the PL intensities of nanoplatelets based on F-PEA ligands are reduced by approximately two orders of magnitude as compared to PEA$_2$EuBr$_4$. Lastly, we compare the absorption, PL, and PLE spectra of L$_2$EuBr$_4$ nanoplatelets in **Figure S8** where all L$_2$EuBr$_4$ nanoplatelets achieve deep blue PL and agreement between respective absorption and PLE spectra.

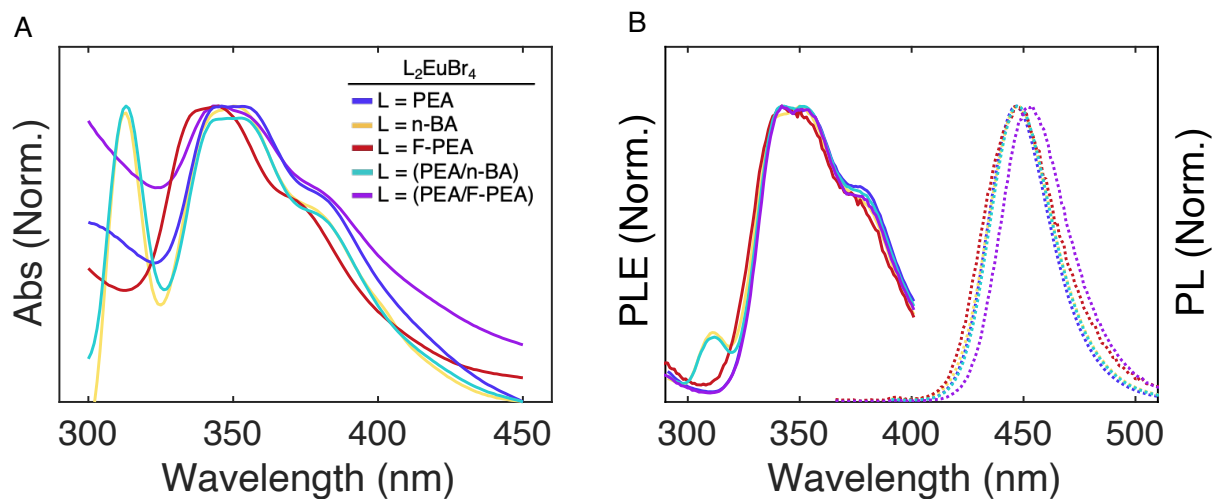

**Figure S8.** Absorption **(A)**, photoluminescence excitation and photoluminescence **(B)** spectra of solution-state L$_2$EuBr$_4$ across various organic ligands.

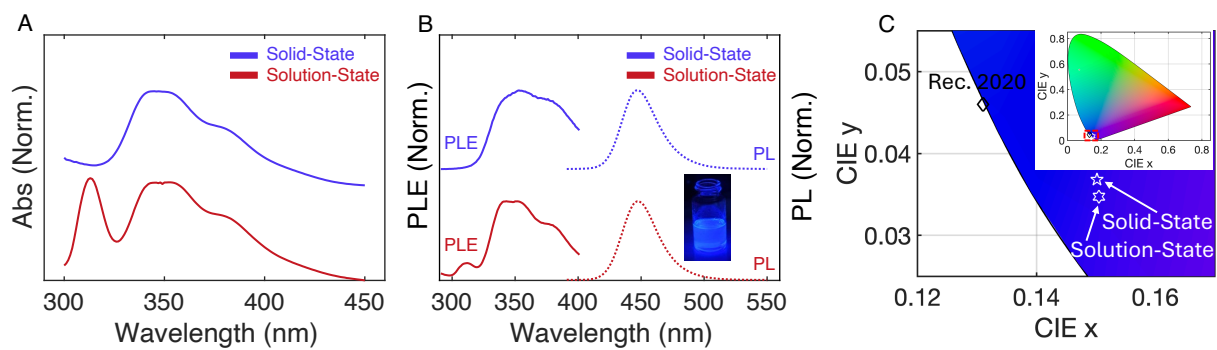

**Figure S9.** Optical properties of (PEA/n-BA)$_2$EuBr$_4$ nanoplatelets. **(A)** Absorption of solid- and solution-state (PEA/n-BA)$_2$EuBr$_4$ nanoplatelets. **(B)** Photoluminescence excitation (left) and photoluminescence (right) spectra for solid- and solution-state (PEA/n-BA)$_2$EuBr$_4$ nanoplatelets. Inset shows solution-state (PEA/n-BA)$_2$EuBr$_4$ nanoplatelets under 365 nm light. **(C)** CIE coordinates for solid- and solution-state (PEA/n-BA)$_2$EuBr$_4$ nanoplatelets. Inset shows the entire CIE chromaticity diagram with the (PEA/n-BA)$_2$EuBr$_4$ nanoplatelets' coordinates outlined in a red dashed box.

Given the comparable PL intensities between (PEA/n-BA)$_2$EuBr$_4$ and PEA$_2$EuBr$_4$ solution-state nanoplatelets, we further analyze the properties of both solid- and solution-state (PEA/n-BA)$_2$EuBr$_4$ nanoplatelets. **Figure S9A** shows the absorption spectra of (PEA/n-BA)$_2$EuBr$_4$ nanoplatelets where both solid- and solution-state (PEA/n-BA)$_2$EuBr$_4$ nanoplatelets show strong absorption at approximately 350 nm, similar to PEA$_2$EuBr$_4$ nanoplatelets. However, solution-state (PEA/n-BA)$_2$EuBr$_4$ nanoplatelets have an additional absorption peak at 313 nm which also appears in the corresponding PLE spectrum in **Figure S9B**. Given the presence of this peak in only the solution-state nanoplatelets, its origins could stem from the interaction between the chlorobenzene solvent and n-butylammonium ligands. **Figure S9B** shows the alignment between the PLE and absorption spectra in **Figure S9A** as well as the PL spectra for both solid- and solution-state (PEA/n-BA)$_2$EuBr$_4$ nanoplatelets. Solid- and solution-state (PEA/n-BA)$_2$EuBr$_4$ nanoplatelets achieve PL emission centered at 446 nm and 448 nm, respectively, with FWHMs of 33 nm. Further, the PLQYs of solid- and solution-state (PEA/n-BA)$_2$EuBr$_4$ nanoplatelets is 7.52% and 8.10%, which equate to decreases of 26.3% and 36.7% compared to PEA$_2$EuBr$_4$, respectively. However, **Figure S9C** shows the CIE color coordinates of both solid- and solution-state (PEA/n-BA)$_2$EuBr$_4$ nanoplatelets [(0.1502, 0.0368) and (0.1505, 0.0347), respectively] which show greater alignment with Rec. 2020's blue CIE color coordinate compared to PEA$_2$EuBr$_4$. The TRPL spectra and decay curve of solid-state (PEA/n-BA)$_2$EuBr$_4$ nanoplatelets is shown in **Figure S10**, where the extracted decay lifetime is 41.8 ns. We then calculate radiative and nonradiative recombination rates of 1.80 x 10$^6$ s$^{-1}$ and 2.21 x 10$^7$ s$^{-1}$, respectively, for solid-state (PEA/n-BA)$_2$EuBr$_4$ nanoplatelets. Compared to PEA$_2$EuBr$_4$, both the radiative and nonradiative recombination rates are increased with a greater relative increase in the nonradiative recombination rate – which yielded a reduced PLQY. We summarize the extracted metrics of (PEA/n-BA)$_2$EuBr$_4$ nanoplatelets in Table S2. Thus, while PEA$_2$EuBr$_4$ nanoplatelets demonstrate greater PL efficiency, (PEA/n-BA)$_2$EuBr$_4$ nanoplatelets show better alignment with Rec. 2020.

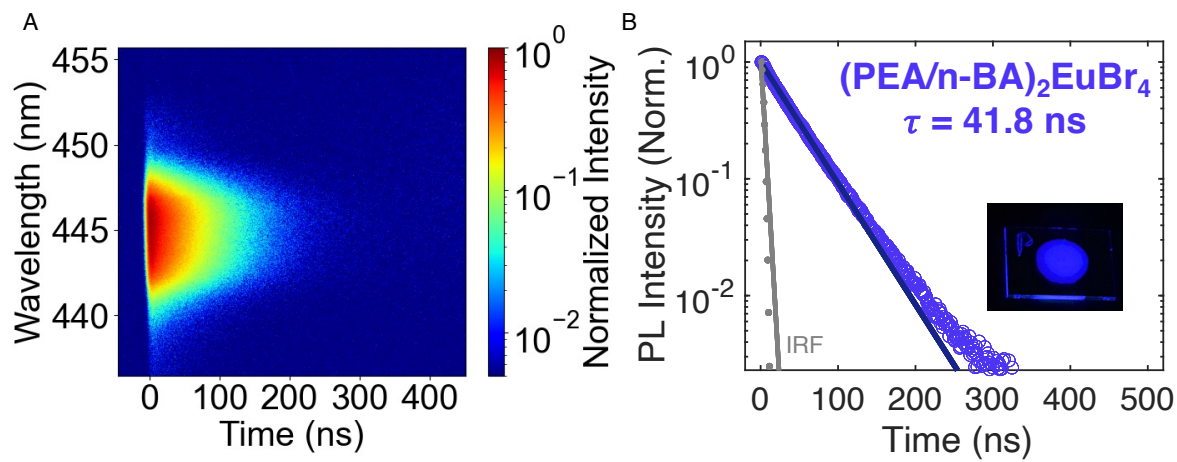

**Figure S10. (A)** Time-resolved photoluminescence spectra of (PEA/n-BA)$_2$EuBr$_4$ nanoplatelets. Note: a 445 nm bandpass filter was used within this optical set-up. **(B)** Time-resolved photoluminescence decay curve measured at 446 nm of solid-state (PEA/n-BA)$_2$EuBr$_4$ nanoplatelets. $R^2$ = 0.999. Inset shows solid-state (PEA/n-BA)$_2$EuBr$_4$ 449nanoplatelets under 365 nm light.

Table S2. Summarized metrics of (PEA/n-BA)$_2$EuBr$_4$ nanoplatelets from **Figures S7** and **S8.**

| (PEA/n-BA)$_2$EuBr$_4$ | PL Peak | FWHM | PLQY | $\tau$ | $k_{rad}$ | $k_{nonrad}$ |
|---|---|---|---|---|---|---|
| Solid-State | 446 nm | 33 nm | 7.52 % | 41.8 ns | 1.80 x 10$^6$ s$^{-1}$ | 2.21 x 10$^7$ s$^{-1}$ |
| Solution-State | 448 nm | 33 nm | 8.10 % | N/A | N/A | N/A |

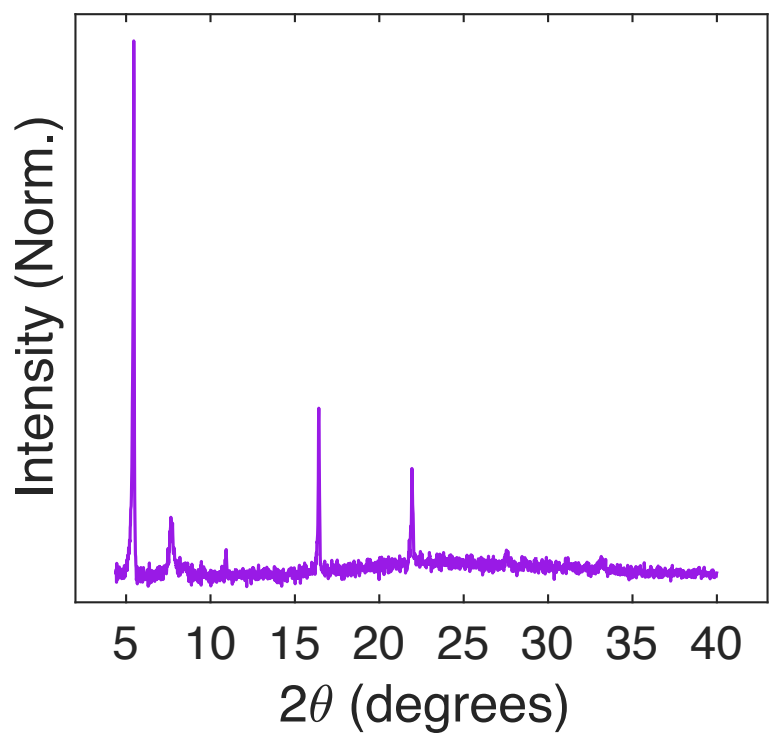

**Figure S11.** XRD Pattern of PEA$_2$EuCl$_4$ nanoplatelets.

Table S3. Periodic XRD peak positions of both PEA$_2$EuBr$_4$ and PEA$_2$EuCl$_4$ nanoplatelets.

| Perovskite | Peak 1 | Peak 2 | Peak 3 | Peak 4 | Peak 5 | Peak 6 | Peak 7 |
|---|---|---|---|---|---|---|---|
| PEA$_2$EuBr$_4$ | 5.51 º | 11.0 º | 16.6 º | 22.2 º | 27.8 º | 33.5 º | N/A |
| PEA$_2$EuCl$_4$ | 5.45 º | 10.9 º | 16.4 º | 21.9 º | 27.5 º | 33.1 º | 7.65 º |

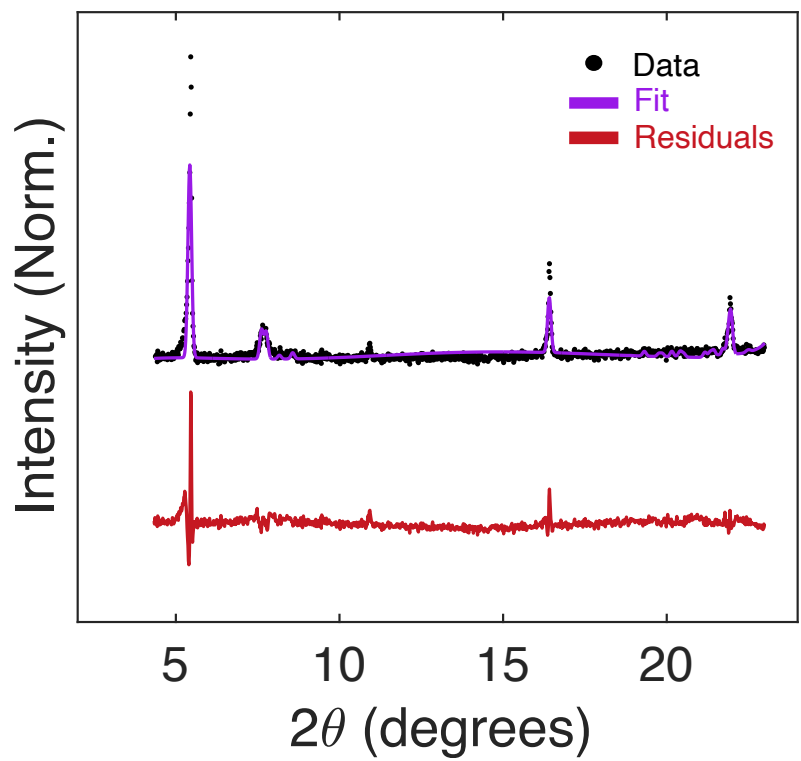

**Figure S12.** Pawley refinement of the XRD pattern for PEA$_2$EuCl$_4$ nanoplatelets.

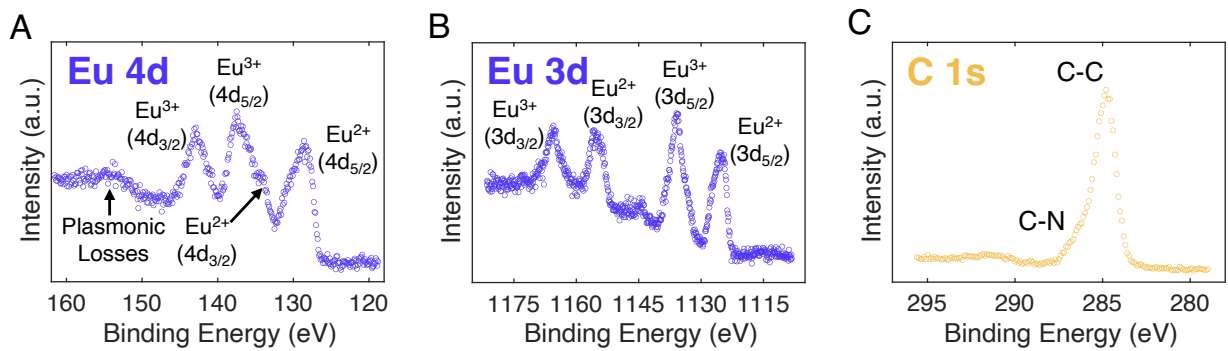

**Figure S13.** XPS spectra of **(A)** Eu 4d, **(B)** Eu 3d, and **(C)** C 1s of PEA$_2$EuCl$_4$ nanoplatelets.

Note: XPS spectra measured without Ar ion beam treatment.

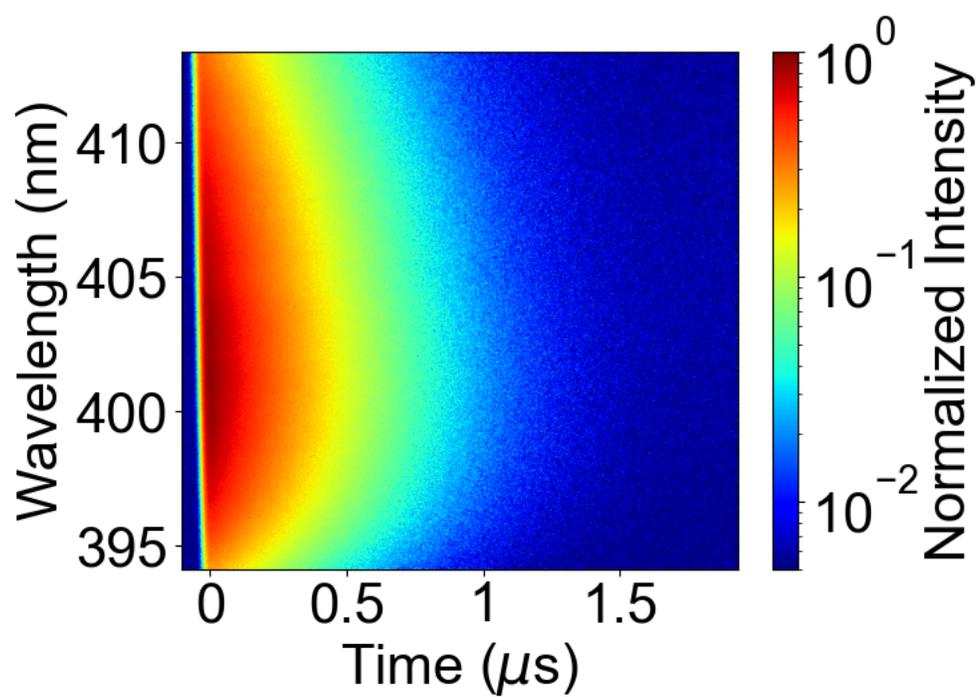

**Figure S14.** Time-resolved photoluminescence spectra of solid-state PEA$_2$EuCl$_4$ nanoplatelets.

Note: a 400 nm longpass filter was used within this optical set-up.